\RequirePackage{letltxmacro}
\LetLtxMacro{\LaTeXtextbf}{\textbf}
\documentclass{ieeeaccess}
\LetLtxMacro{\textbf}{\LaTeXtextbf}
\usepackage{hyperref}
\usepackage{cite}
\usepackage{amsmath,amssymb,amsfonts}
\usepackage{algorithmic}
\usepackage{graphicx}
\usepackage{textcomp}
\def\BibTeX{{\rm B\kern-.05em{\sc i\kern-.025em b}\kern-.08em T\kern-.1667em\lower.7ex\hbox{E}\kern-.125emX}}
\usepackage{enumitem,lipsum}
\usepackage{etoolbox}
\AtBeginEnvironment{itemize}{\setlength\parskip{0pt}}
\AfterEndEnvironment{itemize}{\vspace*{-\dimexpr\parskip\relax}}
\setlist[itemize]{parsep=0pt}
\usepackage{breqn}
\DeclareMathOperator{\E}{\mathbb{E}}
\usepackage[table]{xcolor}
\usepackage{multirow, multicol}
\usepackage{booktabs, makecell}
\usepackage[ruled]{algorithm2e}
\newlength{\commentWidth}
\usepackage{mathtools}

\let\norm\undefined
\DeclarePairedDelimiter\norm{\lVert}{\rVert}
\makeatletter
\newcommand{\mathleft}{\@fleqntrue\@mathmargin0pt}
\newcommand{\mathcenter}{\@fleqnfalse}
\makeatother
\makeatletter
\newcommand*{\rom}[1]{\expandafter\@slowromancap\romannumeral #1@}
\makeatother
\usepackage{caption}
\usepackage{subcaption}
\usepackage{color}
\usepackage{colortbl}
\definecolor{myblue}{RGB}{240,247,255}
\definecolor{mygray}{RGB}{242,242,242}
\definecolor{mygreen}{RGB}{248,252,246}
\setlist[itemize]{align=parleft,left=0pt..1em}
\begin{document}
\history{Date of publication xxxx 00, 0000, date of current version xxxx 00, 0000.}
\doi{10.1109/ACCESS.2017.DOI}
\title{Bootstrap Equilibrium and Probabilistic Speaker Representation Learning for Self-supervised Speaker Verification}
\author{\uppercase{Sung Hwan Mun}, \IEEEmembership{Student Member, IEEE},
\uppercase{Min Hyun Han}, \IEEEmembership{Student Member, IEEE},
\uppercase{Dongjune Lee}, \IEEEmembership{Student Member, IEEE},
\uppercase{Jihwan Kim}, \IEEEmembership{Student Member, IEEE},
\uppercase{and Nam Soo Kim}$^{\dag}$ \IEEEmembership{Senior Member, IEEE}}
\address[]{Department of Electrical and Computer Engineering and INMC, Seoul National University, Seoul 08826, South Korea \\(e-mail: shmun@hi.snu.ac.kr; mhhan@hi.snu.ac.kr; djlee@hi.snu.ac.kr; jhkim21@hi.snu.ac.kr; nkim@snu.ac.kr)}
\tfootnote{This research was supported and funded by the Korean National Police Agency. [Project Name: Real-time speaker recognition via voiceprint analysis / Project Number: PR01-02-040-17]}
\markboth
{Author \headeretal: Preparation of Papers for IEEE TRANSACTIONS and JOURNALS}
{Author \headeretal: Preparation of Papers for IEEE TRANSACTIONS and JOURNALS}
\corresp{$^{\dag}$Corresponding author: Nam Soo Kim (e-mail: nkim@snu.ac.kr).}

\begin{abstract}
In this paper, we propose self-supervised speaker representation learning strategies, which comprise of a \textit{bootstrap equilibrium speaker representation learning} in the front-end and an \textit{uncertainty-aware probabilistic speaker embedding training} in the back-end. In the front-end stage, we learn the speaker representations via the bootstrap training scheme with the uniformity regularization term. In the back-end stage, the probabilistic speaker embeddings are estimated by maximizing the mutual likelihood score between the speech samples belonging to the same speaker, which provide not only speaker representations but also data uncertainty. Experimental results show that the proposed bootstrap equilibrium training strategy can effectively help learn the speaker representations and outperforms the conventional methods based on contrastive learning. Also, we demonstrate that the integrated two-stage framework further improves the speaker verification performance on the VoxCeleb1 test set in terms of EER and MinDCF.
\end{abstract}

\begin{keywords}
Speaker verification, self-supervised learning, bootstrap representation learning, probabilistic speaker embedding.
\end{keywords}

\titlepgskip=-15pt
\maketitle
\section{Introduction}
\label{sec:introduction}
\PARstart{S}{peaker} verification (SV) is the task of verifying whether the claimed identity from given the speech samples is true or not. SV has become one of the key technologies for authentication in e-commerce applications, general business interactions, and forensics \cite{15Hansen}. Generally, the SV system consists of two stages: front-end encoding and back-end scoring. In the front-end encoding stage, the utterance-level fixed-dimensional feature vector is extracted by summarizing speech samples with varying frame lengths. The speaker's voice patterns and characteristics can be condensed in this feature vector. In the back-end scoring stage, the similarity or likelihood (e.g., cosine similarity and probabilistic linear discriminant analysis) between the feature vectors from the enrollment and test utterances is measured to decide the acceptance or rejection of the claimed identity.

In recent years, inspired by the success in various fields such as vision and natural language processing, various deep learning-based methods for the speaker verification have been proposed \cite{18Snyder, 18Wan, 18Okabe, 20Chung, 19Jung}. Among them, the most popular approach is to use utterance-level bottleneck features called speaker embeddings. These deep speaker embedding techniques have significantly boosted the performance when a large amount of training data is available \cite{17Nagrani, 18Chung, 20Nagrani}. However, most of these methods are based on fully supervised learning techniques, which requires a huge amount of manually labeled data. Furthermore, the construction of labeled corpus in real-world scenarios is labour-intensive and sometimes limited by privacy issues.

Self-supervised learning (SSL) is a promising alternative approach that reduces the need for labeling burden. SSL is a branch of unsupervised learning, and it utilizes input data itself as the target for supervision \cite{21Le-Khac, 21Liu}. Recently, the most prevalent methods in SSL are based on contrastive learning \cite{18Oord, 20Chen, 20He}. The core idea in these methods is to pull together two representations jointly sampled from the same class (i.e., positive pairs) while pushing apart those independently sampled from different classes (i.e., negative pairs) through \textit{contrastive loss} such as the InfoNCE \cite{18Oord} or NT-Xent \cite{20Chen}. Also, it is proven that minimizing the \textit{contrastive loss} is equivalent to maximizing the lower bound of the mutual information between latent representations \cite{18Oord, 19Arora}.

Over the last couple of years, several works have been attempted for self-supervised speaker verification. Stafylakis \textit{et al.} \cite{19Stafylakis} exploited a pretext task reconstructing the frames of a target speech segment, given a speaker embedding inferred from another part of the same utterance. Authors in \cite{19Ravanelli, 19Jati} learned speech representations that capture the speaker's identity by maximizing the mutual information between the local embeddings extracted from the same utterance. Furthermore, to generate more robust positive pairs, data augmentation with different additive noises and simulated room impulse responses (RIR) was applied in \cite{20Inoue, 20Huh, 21Zhang, 21Xia}. Huh \textit{et al.} \cite{20Huh} proposed an augmentation adversarial training strategy that penalizes the ability to predict the augmentation types so that the embeddings can be optimized to be channel-invariant. Instead of adversarial training, Zhang \textit{et al.} \cite{21Zhang} adopted a joint training approach using a channel-invariant loss formulated as the distance between the embedding of the augmented segments and its clean version. Xia \textit{et al.} \cite{21Xia} utilized a queue to maintain many negative pairs as in MoCo \cite{20He} and proposed a prototypical memory bank to compensate for the samples wrongly verified as negative.

Although the aforementioned methods have shown successful results, the contrastive learning techniques generally require careful handling of the negative pairs, e.g., large batch sizes \cite{20Chen}, additional memory banks \cite{20He, 21Xia}, etc. \cite{21Le-Khac, 21Liu, 19Arora, 20Tian, 21Robinson}. Moreover, the framework of contrastive learning assumes that every utterance in a mini-batch (or memory bank) contains only one speaker’s speech. This could lead to a class collision problem \cite{19Arora}, \cite{21Xia}; the actual positive sample may be misclassified as a negative. In addition, their performance greatly depends on the augmentation strategies \cite{20Chen}.

To reduce the dependency on how to choose the negative samples, we introduce a \textit{bootstrap} mechanism for learning speaker representations. The bootstrap approach has shown meaningful outcomes in the fields of reinforcement \cite{20Guo}, vision \cite{20Grill, 20Chen2}, graph \cite{21Thakoor, 21Che}, sentence \cite{21Zhang2} and user-item representation learning \cite{21Lee}. They learned the representations by predicting the target latent embeddings of a positive pair, where the asymmetric prediction tasks make a bootstrapping effect in the latent space \cite{20Guo, 20Grill}. In our work, the speaker representations are learned via two distinct networks, namely the \textit{online} and \textit{target} networks. The parameters of both networks are asymmetrically updated with positive pairs. The online network is optimized by back-propagated gradients to predict the outputs of the target network, while the target network is updated through an exponential moving average (EMA) of the online network weights.

The algorithmic components for the bootstrap framework, such as the stop-gradient, EMA, and the asymmetric update, help prevent the representations of all samples from being similar, which is known as the problem of collapsed solutions \cite{20Grill, 20Chen2}.
Nevertheless, we empirically found that learning the speaker embeddings with only a bootstrap prediction task is insufficient to avoid collapsed representations (shown in Section \rom{4}-B.1).
In order to mitigate this difficulty, we increase the entropy of the nuisance factors via a uniformity regularization term.
Optimizing the uniformity regularization term forces the embeddings to be in the equilibrium state (i.e., uniformly distributed over the unit hypersphere), which leads to a maximum entropy in the latent space \cite{06Vignat}.
This can improve the inter-class separability between speaker representations, which was not considered in the previous bootstrap techniques.
To formulate the uniformity regularization loss, we exploit the total pairwise potential based on a Gaussian kernel function, which is closely related to the universally optimal point configurations \cite{07Cohn, 19Borodachov, 20Wang}.

On top of that, we leverage a mutual likelihood score based on an \textit{uncertainty-aware probabilistic speaker embedding} for the back-end stage.
Probabilistic embedding has been introduced in natural language processing \cite{15Vilnis}, computer vision \cite{19Oh, 19Shi, 21Chen}, and other areas \cite{18Bojchevski} to represent the feature (mean) and uncertainty (variance) simultaneously.
Instead of a deterministic point embedding, we estimate the distribution of each speaker embedding, i.e., a Gaussian distribution, where the mean represents the \textit{identity-salient} features, while the variance explains \textit{data uncertainty}.
We make use of the bootstrap speaker representations for the mean vectors and learn the covariance matrices to quantify the data uncertainty in the self-supervised learning fashion.
To optimize the back-end system, we maximize the mutual likelihood score (MLS) between the probabilistic speaker embeddings. After training, the MLS is used for verification.

Experimental results demonstrated that learning the speaker representations via the bootstrap prediction loss and the uniformity regularization loss could avoid collapsing to a trivial solution and further improve the speaker verification performance.
Additionally, we investigated that this bootstrap speaker representation can be more resilient to the batch size compared to its contrastive counterpart.
Finally, by using the MLS between the estimated probabilistic speaker embeddings in the back-end, the proposed framework outperformed the existing self-supervised speaker verification methods in terms of the equal error rate (EER) and minimum detection cost function (MinDCF) on the VoxCeleb1 test set.
The contributions of this paper are as follows:

\begin{itemize}[nosep]
\item We propose the bootstrap training strategy to learn the speaker representations in a self-supervised manner.
The front-end networks are trained via the objective to predict the target embeddings of the positive pairs through the asymmetrical updates.
Furthermore, we introduce the uniformity regularization term to prevent the speaker representations from collapsing into a trivial solution.
This regularization term increases the entropy on nuisance factors inherent in the speaker embeddings, which can lead to the enhancement of the inter-speaker separability.
By minimizing the combined loss, we learn the \textit{bootstrap equilibrium speaker representation} in the front-end stage.
\item We incorporate the data uncertainty in the verification step of the back-end scoring stage. To represent the data uncertainty on the input speech, we introduce the \textit{uncertainty-aware probabilistic speaker embedding} approach. Unlike the conventional deterministic point embedding, the probabilistic speaker embedding follows the Gaussian distribution where the mean and variance provide the speaker identity and data uncertainty, respectively.
The parameters of the probabilistic speaker embeddings are estimated by maximizing the MLS.
\item
Experiments investigate the training analysis and the speaker verification performance in terms of EER and MinDCF on the VoxCeleb 1 evaluation set. Also, we compare the proposed methods with the conventional techniques. By integrating the two proposed stages, we achieved outstanding results in the self-supervised speaker verification task, outperforming the existing techniques.
\end{itemize}

\vspace{1.0mm}
The rest of this paper is organized as follows: In Section \rom{2}, we introduce the conventional self-supervised speaker verification framework. In Section \rom{3}, the proposed training strategies are described. The experimental results are shown in Section \rom{4}. Finally, Section \rom{5} concludes the paper.

%%%%%%%%%%%%%%%%%%%%%%%%%%%%%%%%%%%%%%%%%%%%%%%%%%%%%%%%%%%%%%%%%%%%%%%%%%%%%%%%%%%%%%%%%%%%%%%%%%%%%%%%%
\section{Conventional self-supervised Contrastive Speaker Representation Learning}
Let $X=\{x_{1}, \dots, x_{N}\}$ represent the $N$-utterances in each mini-batch $\mathcal{B}$ randomly sampled from an unlabeled training dataset.
In the contrastive speaker representation learning framework shown in FIGURE 1, to obtain two segments from each utterance $x_i$, the non-overlapping segments $x_{i,1}$ and $x_{i,2}$ are randomly cropped with the same frame length $T$.
Under the assumption that every utterance within a mini-batch $\mathcal{B}$ has a single speaker's identity, $(x_{i,1},x_{i,2})$ denotes a positive pair, i.e., having the same speaker identity.

A data augmentation policy is then randomly sampled for each of the segments $x_{i,1}$ and $x_{i,2}$, e.g., $(\mathcal{N}', \mathcal{R}') \sim \mathcal{T}$ and $(\mathcal{N}'', \mathcal{R}'') \sim \mathcal{T}$, as follows:
\begin{equation}
\begin{split}
   y'_{\vartheta;i,1} &= f_{\vartheta}(x'_{i,1}) = f_{\vartheta}(x_{i,1} * \mathcal{R}' + \mathcal{N}'), \\
   y''_{\vartheta;i,2} &= f_{\vartheta}(x''_{i,2}) = f_{\vartheta}(x_{i,2} * \mathcal{R}'' + \mathcal{N}''),
\end{split}
\end{equation}
where $\mathcal{N}'$, $\mathcal{N}''$ and $\mathcal{R}'$, $\mathcal{R}''$ are random noises and RIR filters, respectively.
$x'_{i,1}$ and $x''_{i,2}$ denote two differently augmented segments from $x_{i,1}$ and $x_{i,2}$.
$f_{\vartheta}:\mathbb{R}^{D\times T} \rightarrow \mathbb{R}^{d}$ is a siamese front-end encoder that maps the speech segments of dimension $D$ with the frame length $T$ to the embeddings of dimension $d$, and notation $*$ is the convolution operator. $y'_{\vartheta; i,1}$ and $y''_{\vartheta; i,2}$ represent the speaker embeddings.

In order to ensure the embeddings of the positive pairs to be similar while pushing those from the negative pairs apart, an angular prototypical (AP) loss function \cite{20Chung} can be used.
AP loss serves as \textit{contrastive loss} and has been shown to perform well in self-supervised speaker verification tasks \cite{20Huh, 21Zhang, 21Xia}, which is defined as follows:
\begin{equation}
  \mathcal{L}_{\vartheta}^{\mathtt{ap}} = - {1 \over N} \sum_{i=1}^{N} \log { e^{\text{S}(y'_{\vartheta;i,1},y''_{\vartheta;i,2})}  \over {\sum_{j=1}^{N} e^{\text{S}(y'_{\vartheta;i,1},y''_{\vartheta;j,2})} } },
\end{equation}
\begin{equation}
  \text{S}(y'_{\vartheta;i,1},y''_{\vartheta;j,2}) = 
  w {{{y'_{\vartheta;i,1}}^{T} \cdot {y''_{\vartheta;j,2}}} \over 
  {\norm{y'_{\vartheta;i,1}}_2 \norm{y''_{\vartheta;j,2}}_2}} + b.
\end{equation}
In the above expressions, $\text{S}(\cdot): \mathbb{R}^{d} \times \mathbb{R}^{d}\rightarrow \mathbb{R}$ is the affine transformation of the cosine similarity between two speaker embeddings of dimension $d$. $w$ and $b$ are trainable parameters for scale and bias, respectively.
% %%%%%%%%%%%%%%%%%%%%%%%%%%%%%%%%%%% FIGURE 1 %%%%%%%%%%%%%%%%%%%%%%%%%%%%%%%%%%%
\begin{figure}[t!]
\begin{minipage}[b]{1.0\linewidth}
  \centering
  \centerline{\includegraphics[width=8.0cm]{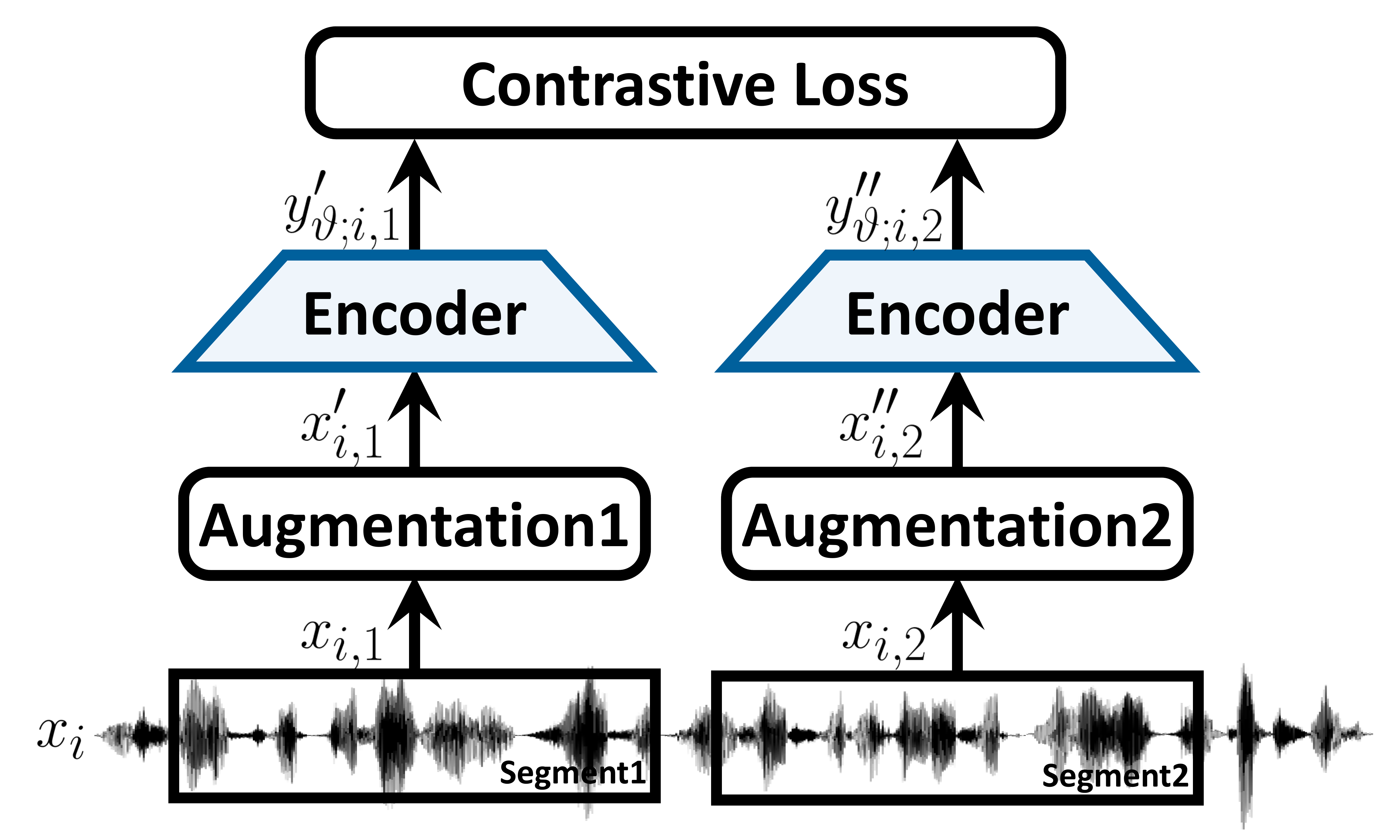}}
\vspace{0cm}
\end{minipage}
\centering
\caption{Self-supervised contrastive speaker representation learning framework.}
\label{figures}
\end{figure}

\section{Proposed Framework}
Our proposed training strategies are composed of two stages: \textit{bootstrap equilibrium speaker representation learning} and \textit{probabilistic speaker embedding training}.
For the front-end stage, we learn the speaker representations via a bootstrap training strategy. And then, in the back-end stage, the distribution of each speaker embedding is estimated by maximizing a mutual likelihood score (MLS).
With the MLS calculated via the estimated distribution, enrollment and test utterances are evaluated.
All methods in our proposed framework operate in a self-supervised learning fashion.

\subsection{Front-end: Bootstrap Equilibrium Speaker Representation Learning}
\subsubsection{Bootstrap training strategy and prediction loss}
Unlike the contrastive methods requiring a careful design of informative negative samples, we learn the speaker representations via a bootstrap mechanism with only positive samples.
Analogous to the previous works \cite{20Guo}, \cite{20Grill}, \cite{21Thakoor}, \cite{21Che}, the proposed framework contains two neural networks: the \textit{online} and the \textit{target} networks.
Let $f:\mathbb{R}^{D\times T} \rightarrow \mathbb{R}^{d_f}$ be an \textit{encoder}, $g:\mathbb{R}^{d_f} \rightarrow \mathbb{R}^{d_g}$ be a \textit{projector}, and $q:\mathbb{R}^{d_g} \rightarrow \mathbb{R}^{d_q}$ be a \textit{predictor}. 
The encoder $f$ maps an input $D\times T$-dimensional speech segment $x$ into a $d_f$-dimensional representation $y \triangleq f(x)$.
The projector $g$ transforms a representation $y$ into a $d_g$-dimensional representation $z \triangleq g(y)$ on the smaller space via non-linear projection layers.
The predictor $q$ outputs a $d_q$-dimensional representation $q(z)$, which is used for a regression task.

The online network is defined by a set of trainable parameters $\theta$ and includes the online encoder $f_\theta$, the online projector $g_\theta$ and the online predictor $q_\theta$.
The target network is parameterized by a set of weights $\xi$ distinct from the online network. It contains the target encoder $f_\xi$ and the target projector $g_\xi$, which have the same architectures as $f_\theta$ and $g_\theta$, respectively.
The predictor network exists only the online branch, which forms an asymmetrical structure between the online and target pipeline (shown in FIGURE 2).

The online parameters $\theta$ are optimized by following the gradients of the euclidean distance between the $\ell_2$-normalized online predictions and target projections. We define a \textit{bootstrap prediction loss} as follows:

\begin{equation}
\begin{split}
  \ell_{\theta,\xi}^{\mathtt{pred}} & \triangleq \displaystyle \E \left[ {\left\lVert \overline{q_\theta}(z'_{\theta}) - {\overline{z}''_{\xi}} \right\rVert}^2_2 \right]\\
  & = {1 \over N}\sum^N_{i=1} \left[ 2 - 2{{q_\theta(z'_{\theta;i,1})}^T \cdot {z''_{\xi;i,2}} \over \norm{{q_\theta(z'_{\theta;i,1})}}_2 \norm{z''_{\xi;i,2}}_2}\right],
\end{split}
\end{equation}

\begin{equation}
  \mathcal{L}_{\theta,\xi}^{\mathtt{pred}}  = \ell_{\theta,\xi}^{\mathtt{pred}} + \tilde{\ell}_{\theta,\xi}^{\mathtt{pred}},
\end{equation}
where $\overline{q_\theta}(z'_{\theta;i,1})$ and ${\overline{z}''_{\xi;i,2}}$ denote the $\ell_2$-normalized online prediction of $x'_{i,1}$ and the $\ell_2$-normalized target projection of $x''_{i,2}$, respectively. 
We also symmetrize the loss of equation (4) as $\tilde{\ell}_{\theta,\xi}^{\mathtt{pred}}$ by predicting the target projection of the input $x'_{i,1}$ using the online prediction of $x''_{i,2}$. Finally, the bootstrap prediction loss $\mathcal{L}_{\theta,\xi}^{\mathtt{pred}}$ is obtained by the sum of two symmetrical losses as in equation (5).

At the training step, the online parameters $\theta$ are updated via back-propagated gradients to minimize $\mathcal{L}_{\theta,\xi}^{\mathtt{pred}}$ with respect to $\theta$ \textit{only}, where the parameters $\xi$ are fixed by the stop-gradient.
The target weights $\xi$ are optimized via the exponential moving average (EMA) of the online parameters $\theta$.
The overall dynamics are as follows: 
\begin{gather}
  \theta \leftarrow \mathtt{optimizer}(\theta, \eta, \nabla_\theta \mathcal{L}_{\theta,\xi}^{\mathtt{pred}}, \xi), \\
  \xi \leftarrow \tau\xi + (1-\tau)\theta,
\end{gather}
where $\mathtt{optimizer}$ is an optimizer such as SGD or Adam, and $\eta$ is the learning rate. $\tau$ $\in [0,1]$ is a target decay rate for the momentum-based exponential moving average.

By gradually increasing a value of $\tau$ for each training step, the target network slowly approximates the online encoder.
This momentum-based update makes $\xi$ develop more smoothly than $\theta$, which allows to \textit{bootstrap} the representations by providing enhanced but consistent targets to the online network \cite{20Grill}.
In our work, we set $\tau \triangleq 1-(1-\tau_{\mathtt{base}})\cdot(\cos(\pi k/K)+1)/2$ where $\tau_{\mathtt{base}}$ is an initial base value, $k$ is the current training steps, and $K$ is the maximum number of training steps. 
The coefficient $\tau$ increases from $\tau_{\mathtt{base}}$ to 1 according to the above schedule during training.

% %%%%%%%%%%%%%%%%%%%%%%%%%%%%%%%%%%% ALGORITHM 1 %%%%%%%%%%%%%%%%%%%%%%%%%%%%%%%%%%%
\begin{algorithm}[t!]
\caption{Bootstrap equilibrium training strategy}
\SetAlgoLined
\small{
\KwIn{$X, \, N, \, K, \mathcal{T}, (\theta; \, f_{\theta}, \, g_{\theta}, \, q_{\theta}), (\xi; \, f_{\xi}, \, g_{\xi}), \break \{ \tau_k \}_{k=1}^{K}, \{ \eta_k \}_{k=1}^{K} $ and $\mathtt{\tiny optimizer}.$}

Init. randomly: $(f_{\theta}, \, g_{\theta}, \, q_{\theta}) \gets \mathtt{Kaiming\,He}(\theta)$ \\
Init. randomly: $(f_{\xi}, \, g_{\xi}) \gets \mathtt{Kaiming\,He}(\xi)$ \\
\For{$k= $ $ 1 $ to $ K$}{
    $\mathcal{B} \gets \{x_i \sim X \}_{i=1}^N$ \\
    \For{$i= $ $ 1 $ to $ N$}{
        $(x_{i,1}, x_{i,2}) \sim x_{i} $ \\
        $(\mathcal{N'}, \mathcal{R'}) \sim \mathcal{T}$, $(\mathcal{N''}, \mathcal{R''}) \sim \mathcal{T}$  \\
        
        $x'_{i,1} \gets x_{i,1}*\mathcal{R'}+\mathcal{N'}$ \\
        $x''_{i,2} \gets x_{i,2}*\mathcal{R''}+\mathcal{N''}$ \\
        $\mathcal{S} \gets \mathtt{append} (x'_{i,1}, x''_{i,2})$ \\
    }
    \For{$i= $ $ 1 $ to $ N$ }{
        $\ell_{\theta,\xi;i}^{\mathtt{pred}} \gets - {{ 2 \cdot {{q_\theta}(g_{\theta}(f_{\theta}(x'_{i,1})))}^T \cdot {g_{\xi}}(f_{\xi}(x''_{i,2})) } \over { \norm{ {q_\theta}(g_{\theta}(f_{\theta}(x'_{i,1})))}_2 {\norm{{g_{\xi}}(f_{\xi}(x''_{i,2}))}}_2}}$ \\
        $\tilde{\ell}_{\theta,\xi;i}^{\mathtt{pred}} \gets - {{ {2 \cdot {q_\theta}(g_{\theta}(f_{\theta}(x''_{i,2})))}^T \cdot {g_{\xi}}(f_{\xi}(x'_{i,1})) } \over { \norm{ {q_\theta}(g_{\theta}(f_{\theta}(x''_{i,2})))}_2 {\norm{{g_{\xi}}(f_{\xi}(x'_{i,1}))}}_2}}$ \\
        $\mathcal{L}_{\theta,\xi;i}^{\mathtt{pred}} \gets \ell_{\theta,\xi;i}^{\mathtt{pred}} + \tilde{\ell}_{\theta,\xi;i}^{\mathtt{pred}} $ \\
    \vspace{0.75mm}
        \For{ $j= $ $ 1 $ to $ N$ }{
            $\ell_{\theta,\xi;i,j}^{\mathtt{unif}} \gets e^{-t {\left\lVert \overline{q_\theta}(g_{\theta}(f_{\theta}(x'_{i,1}))) - {\overline{g_{\xi}}(f_{\xi}(x''_{j,2})}  \right\rVert}_2^2}$ \\
            $\tilde{\ell}_{\theta,\xi;i,j}^{\mathtt{unif}} \gets e^{-t {\left\lVert \overline{q_\theta}(g_{\theta}(f_{\theta}(x''_{j,2}))) - {\overline{g_{\xi}}(f_{\xi}(x'_{i,1})}  \right\rVert}_2^2}$ \\
            $\mathcal{L}_{\theta,\xi;i,j}^{\mathtt{unif}} \gets \ell_{\theta,\xi;i,j}^{\mathtt{unif}} + \tilde{\ell}_{\theta,\xi;i,j}^{\mathtt{unif}} $
        }
    }
    $\nabla_{\theta}^{\mathtt{pred}} \gets \nabla_\theta {1 \over N} \sum_{i}{\mathcal{L}_{\theta,\xi;i}^{\mathtt{pred}}} $ \\
    $\nabla_{\theta}^{\mathtt{unif}} \gets \nabla_\theta \log{{1 \over N^2} \sum_{i,j}\mathcal{L}_{\theta,\xi;i,j}^{\mathtt{unif}} } $ \\
    $\theta \gets \mathtt{\tiny optimizer}(\theta, \eta_k, \nabla_{\theta}^{\mathtt{pred}}+\nabla_{\theta}^{\mathtt{unif}}) $ \\
    $\xi \leftarrow \tau_k \xi + (1-\tau_k)\theta$
}
\KwOut{online encoder $f_\theta$.}
}
\end{algorithm}

\subsubsection{Uniformity regularization loss}
To force the embeddings to reach an equilibrium state, namely a state of minimal energy (i.e., the distribution optimizing this metric should converge to the uniform distribution on the hypersphere), we leverage the pairwise Gaussian potential kernel as follows:
\begin{equation}
\begin{split}
  \text{G}_t(h(x_i), h(x_j)) \triangleq 
  e^{-t {\left\lVert h(x_i)- h(x_j) \right\rVert}^2_2},
\end{split}
\end{equation}
where $h: \mathbb{R}^{D \times T} \rightarrow \mathbb{R}^d $ is an encoder function, and $t > 0$ is a fixed parameter. Similar to \cite{20Wang, 07Cohn, 19Borodachov}, the uniformity regularization loss is defined as the logarithm of the average pairwise Gaussian potential as follows:
\begin{equation}
\begin{split}
  \ell & \triangleq \log \displaystyle \E_{x_i, x_j \overset{\text{\tiny{i.i.d.}}}{\sim}  p_{\texttt{\tiny data}}} \left[\text{G}_t(h(x_i), h(x_j))\right] \\
  & = \log \displaystyle \E_{x_i, x_j \overset{\text{\tiny{i.i.d.}}}{\sim}  p_{\texttt{\tiny data}}} \left[e^{-t {\left\lVert h(x_i) - h(x_j) \right\rVert}^2_2} \right],
\end{split}
\end{equation}
where minimizing equation (9) leads the embedding vectors to have on uniform distribution \cite{06Vignat, 07Cohn}.
We apply the uniformity regularization loss to the predictions and projections of all pairs as follows:
\begin{equation}
\begin{split}
  \ell_{\theta,\xi}^{\mathtt{unif}} & \triangleq \log \displaystyle \E_{x'_i, x''_j \in \mathcal{B}} \left[ \text{G}_t( \overline{q_\theta}(z'_{\theta}), {\overline{z}''_{\xi}} ) \right] \\
  & = \log \displaystyle \E_{x'_i, x''_j \in \mathcal{B}} \left[e^{-t {\left\lVert \overline{q_\theta}(z'_{\theta}) - {\overline{z}''_{\xi}} \right\rVert}^2_2 }\right].
\end{split}
\end{equation}
In practice, the uniformity regularization loss within the mini-batch can be calculated as follows:
\begin{equation}
   \ell_{\theta,\xi}^{\mathtt{unif}} = \log {{1 \over N^2} \sum_{i=1}^N\sum_{j=1}^N{ e^{-t {\left\lVert \overline{q_\theta}(z'_{\theta;i,1}) - {\overline{z}''_{\xi;j,2}} \right\rVert}^2_2}}} ,\\
\end{equation}
\begin{equation}
   \mathcal{L}_{\theta,\xi}^{\mathtt{unif}} = \ell_{\theta,\xi}^{\mathtt{unif}} + \tilde{\ell}_{\theta,\xi}^{\mathtt{unif}}.
\end{equation}
As in equation (5) of the bootstrap prediction loss, the symmetrical form of equation (11), i.e., $\tilde{\ell}_{\theta,\xi}^{\mathtt{unif}}$, is computed by putting the input $x'_{i,1}$ into the target network and $x''_{j,2}$ into the online network. Total uniformity regularization loss is the sum of the two symmetrical losses.

\subsubsection{Bootstrapped equilibrium speaker embedding}
For the front-end stage, the online and target networks are trained using both the bootstrap prediction loss and the uniformity regularization loss. The total objective and dynamics of the front-end are as follows:
\begin{gather}
  \mathcal{L}_{\theta,\xi}^{\mathtt{front}} = \mathcal{L}_{\theta,\xi}^{\mathtt{pred}} + \mathcal{L}_{\theta,\xi}^{\mathtt{unif}}, \\
  \theta \leftarrow \mathtt{optimizer}(\theta, \eta, \nabla_\theta \mathcal{L}_{\theta,\xi}^{\mathtt{front}}, \xi), \\
  \xi \leftarrow \tau\xi + (1-\tau)\theta.
\end{gather}
After the training, we only keep the online encoder $f_\theta$ and use it as the front-end encoder.
The representations $y_\theta \triangleq f_\theta(x)$ are extracted as the speaker embeddings.
Algorithm 1 summarizes the bootstrap equilibrium training strategy.

\subsection{Back-end: Uncertainty-aware probabilistic speaker embedding training}
\subsubsection{Probabilistic speaker embedding and MLS}
The speaker representation in the front-end is learned as a deterministic point embedding for each speech segment.
In contrast to the conventional deterministic speaker embedding, we estimate the distribution of each speaker embedding in the back-end stage.
In several fields \cite{15Vilnis, 18Bojchevski, 19Oh, 19Shi, 21Chen}, there have been trials using probabilistic representations to estimate data uncertainty.
Similar to \cite{19Shi, 21Chen}, we model each speaker embedding as the Gaussian distribution as follows:
\begin{equation}
  p(y|x_i) = \mathcal{N}(y; \mu_i, \sigma_i^{2}\text{I}),
\end{equation}
where $\mu_i$ and $\sigma_i^2$ are the $d$-dimensional mean and variance vectors. We consider the diagonal covariance matrix to reduce the complexity. Given the probabilistic speaker embeddings of two speech segments, the \textit{mutual likelihood score} (MLS) can be measured as follows:
\begin{equation}
  p(y_i=y_j) = \int p(y_i|x_i)p(y_j|x_j)\delta(y_i-y_j)dy_idy_j,
\end{equation}
where $y_i \sim \mathcal{N}(y; \mu_i, \sigma_i^{2}\text{I})$ and $y_j \sim \mathcal{N}(y; \mu_j, \sigma_j^{2}\text{I})$. The MLS expresses the \textit{likelihood} of two representations belonging to the same speaker [11]. By taking the log-likelihood form, MLS can be formulated as follows:
\begin{align}
  \text{mls}(x_i, x_j) & = \log p(y_i=y_j) \nonumber \\
  & = - {1 \over 2} \sum_{l=1}^{d}{ ( {{(\mu_i^{(l)} - \mu_j^{(l)})^2}\over{ \sigma_i^{2(l)} + \sigma_j^{2(l)} }}   + \log(\sigma_i^{2(l)} + \sigma_j^{2(l)}) ) } \nonumber \\
  & \;\;\;\; - const,
\end{align}
where $const={d\over2}\log2\pi$. $\mu_i^{(l)}$ and $\sigma_i^{(l)}$ denote the $l^{th}$ dimension element of $\mu_i$ and $\sigma_i$, respectively. $\mu_i$ of the probabilistic speaker embedding represents the \textit{identity-salient} features, while $\sigma_i$ explains the \textit{data uncertainty}.

For the mean of the probabilistic speaker embedding, we leverage the \textit{bootstrap equilibrium speaker representation} of the front-end stage, i.e., $\mu_i \triangleq f_\theta(x_i)$.
With its parameters $\theta$ fixed, we train the auxiliary uncertainty estimator network to quantify the data uncertainty using the MLS loss.
MLS and MLS loss are formulated as follows:
\begin{align}
  \text{mls}_{\varphi, \theta}(x'_{i,1}&, x''_{i,2}) \nonumber \\
  = - {1 \over 2} &\sum_{l=1}^{d} \Big( {{( f^{(l)}_\theta(x'_{i,1}) - f^{(l)}_\theta(x''_{i,2}))^2}\over{ u_{\varphi}^{(l)}(m_\theta(x'_{i,1})) + u_{\varphi}^{(l)}(m_\theta(x''_{i,2})) }} \nonumber \\
  + \log(&u_{\varphi}^{(l)}(m_\theta(x'_{i,1})) + u_{\varphi}^{(l)}(m_\theta(x''_{i,2}))) \Big),
\end{align}
\begin{equation}
   \mathcal{L}_{\varphi, \theta}^{\mathtt{mls}} = {1 \over N} \sum_{i=1}^N - \text{mls}_{\varphi, \theta}(x'_{i,1}, x''_{i,2}),
\end{equation}
where $m_\theta:\mathbb{R}^{D\times T} \rightarrow \mathbb{R}^{d_m}$ is the multi-level representation fusion module with fixed weights $\theta$, and $u_{\varphi}:\mathbb{R}^{d_m} \rightarrow \mathbb{R}^{d}$ is the uncertainty estimator with trainable parameters $\varphi$.
In the module $m_\theta$, the intermediate layers of the online encoder $f_\theta$ are concatenated into the $d_m$-dimensional multi-level representation for including different level features of the input speech.
The mean vector of the probabilistic speaker embedding is obtained by feeding the input segment into the fixed front-end encoder $f_\theta$ (shown in FIGURE 2).

% %%%%%%%%%%%%%%%%%%%%%%%%%%%%%%%%%%% ALGORITHM 2 %%%%%%%%%%%%%%%%%%%%%%%%%%%%%%%%%%%
\begin{algorithm}[t!]
\caption{Probabilistic speaker embedding}
\SetAlgoLined
\small{
\KwIn{$X, \, N, \, K, \mathcal{T}, \, (\theta; \, f_{\theta}, m_{\theta}), (\varphi; \, u_{\varphi}), \break \{ \eta_k \}_{k=1}^{K} $ and $\mathtt{\tiny optimizer}.$}
%\vspace{2.0mm}
Init. randomly: $u_{\varphi} \gets \mathtt{Kaiming\,He}(\varphi)$ \\
Init. with pre-trained $\theta$: $f_{\theta} \gets f^{\mathtt{front}}_{\theta}$ \\
\For{$k= $ $ 1 $ to $ K$}{
    $\mathcal{B} \gets \{x_i \sim X \}_{i=1}^N$ \\
    \For{$i= $ $ 1 $ to $ N$}{
        $(x_{i,1}, x_{i,2}) \sim x_{i} $ \\
        $(\mathcal{N'}, \mathcal{R'}) \sim \mathcal{T}$, $(\mathcal{N''}, \mathcal{R''}) \sim \mathcal{T}$  \\
        $x'_{i,1} \gets x_{i,1}*\mathcal{R'}+\mathcal{N'}$ \\
        $x''_{i,2} \gets x_{i,2}*\mathcal{R''}+\mathcal{N''}$ \\
        $\rho'_{i} \gets  u_{\varphi}(m_{\theta}(x'_{i,1})) \oslash (u_{\varphi, \mathtt{avg}}(m_{\theta}(x'_{1})) + \epsilon)$ \\ 
        $\rho''_{i} \gets  u_{\varphi}(m_{\theta}(x''_{i,2})) \oslash (u_{\varphi, \mathtt{avg}}(m_{\theta}(x''_{2})) + \epsilon) $ \\ 
        \vspace{2.0mm}
        $\mathcal{L}_{\varphi,\theta;i}^{\mathtt{mls}} \gets -\text{mls}_{\varphi, \theta}(x'_{i,1}, x''_{i,2})$ \\
        %\vspace{1.5mm}
        $\mathcal{L}_{\varphi,\theta;i}^{\mathtt{cnst}} \gets \norm{1-\rho'_i}_2^2 + \norm{1-\rho''_i}_2^2 $ \\
    }
    $\nabla_{\varphi}^{\mathtt{mls}} \gets \nabla_\varphi {1 \over N} \sum_{i}{\mathcal{L}_{\varphi,\theta;i}^{\mathtt{mls}}} $ \\
    $\nabla_{\varphi}^{\mathtt{cnst}} \gets \nabla_\varphi {1 \over N} \sum_{i}{\mathcal{L}_{\varphi,\theta;i}^{\mathtt{cnst}}} $ \\
    $\varphi \gets \mathtt{\tiny optimizer}(\varphi, \eta_k, \nabla_{\varphi}^{\mathtt{mls}}+\nabla_{\varphi}^{\mathtt{cnst}}) $ \\
}
\KwOut{uncertainty estimator $u_\varphi$.}
}
\vspace{2.0mm}
\end{algorithm}

\subsubsection{Uncertainty constraint loss}
To prevent the network from being over-confident in estimating the data uncertainty, we constraint the dynamic range of estimated uncertainty via the following \textit{uncertainty constraint loss}:
\begin{equation}
  \ell_{\varphi, \theta}^{\mathtt{cnst}}(x'_{i,1}) = {1 \over N} \sum_{i=1}^N \sum_{l=1}^d \left( 1 - {u_{\varphi}^{(l)}(m_\theta(x'_{i,1})) \over u_{\varphi,\mathtt{avg}}^{(l)}(m_\theta(x'_{1}))} \right)^2,
\end{equation}
\begin{equation}
   \mathcal{L}_{\varphi, \theta}^{\mathtt{cnst}} = \ell_{\varphi, \theta}^{\mathtt{cnst}}(x'_{i,1}) + \ell_{\varphi, \theta}^{\mathtt{cnst}}(x''_{i,2}),
\end{equation}
where $u_{\varphi,\mathtt{avg}}(m_\theta(x'_{1}))$ is equal to ${1\over N}\sum_{i} u_{\varphi}(m_\theta(x'_{i,1}))$ and $\ell^{\mathtt{cnst}}_{\varphi, \theta}(x''_{i,2})$ is the formulation of the segments $x''_{i,2}$.
This constraint ensures that the uncertainty does not deviate too far from its average, making the estimated uncertainty more reasonable \cite{21Chen}.

\subsubsection{Uncertainty-aware probabilistic speaker embedding}
For the back-end stage, the uncertainty estimator network is optimized via the following total objective and dynamics:
\begin{gather}
  \mathcal{L}_{\varphi, \theta}^{\mathtt{back}} = \mathcal{L}_{\varphi, \theta}^{\mathtt{mls}} + \mathcal{L}_{\varphi, \theta}^{\mathtt{cnst}}, \\
  \varphi \leftarrow \mathtt{optimizer}(\varphi, \eta, \nabla_\varphi \mathcal{L}_{\varphi,\theta}^{\mathtt{back}}, \theta).
\end{gather}
In the evaluation, MLS is measured for the verification score between the trial utterances.
Algorithm 2 sums up the training process, where $\oslash$ refers to an element-wise divide operator.

\section{Experiments}
\subsection{Experimental settings}
\subsubsection{Datasets}
To evaluate the performance of the proposed speaker representation learning framework, we conducted experiments based on the VoxCeleb1 and VoxCeleb2 datasets \cite{17Nagrani, 18Chung, 20Nagrani}.
VoxCeleb dataset is one of the most popular corpus for large-scale text-independent speaker verification and composed of development and test sets with no overlapping speakers. The speech samples were extracted from YouTube video clips, degraded with real-world noises, including background chatter, laughter, overlapping speech, room acoustics, etc.
For learning the proposed speaker representations, we used the development sets of VoxCeleb1 and VoxCeleb2, which consist of 1,092,009 and 148,642 utterances from 5,994 and 1,211 speakers, respectively.
All networks were trained in a fully self-supervised learning manner without using any speaker labels.
The evaluation was performed on the test set of VoxCeleb1, which is composed of 4,874 utterances spoken by 40 speakers. We followed the original VoxCeleb1 test list, including 37,720 trial pairs.

\subsubsection{Model architectures}
The proposed framework comprises three networks: the online, the target, and the uncertainty networks. In the front-end stage, the online and target networks were used for learning the bootstrap equilibrium speaker representations. They were asymmetrically updated via the Adam optimizer and EMA respectively.
For the back-end stage, the online and uncertainty networks were leveraged to estimate the probabilistic speaker embeddings (shown in FIGURE 2).
The detailed configurations of the networks are as follows:

% %%%%%%%%%%%%%%%%%%%%%%%%%%%%%%%%%%% TABLE 1 %%%%%%%%%%%%%%%%%%%%%%%%%%%%%%%%%%%
\begin{table}[t!]
  \caption{Fast ResNet34 network for the front-end encoder.}
  \centering
  \renewcommand{\arraystretch}{1.05}
  \renewcommand{\tabcolsep}{1.3mm}
  {
  \small{
  \begin{tabular}{c|c|c}
    \specialrule{.1em}{.0em}{.0em}
    \multicolumn{1}{c|}{\textbf{Layer}} & \textbf{Fast ResNet34} & \textbf{Output Size} \\
    \specialrule{.07em}{.0em}{.0em}
    \multicolumn{1}{c|}{Input} & log Mel-filterbanks & $40 \times T \times 1$ \\

    \specialrule{.05em}{.0em}{.0em}
    \multicolumn{1}{c|}{\multirow{2}{*}{Conv1}} & $7 \times 7, 16, \text{stride}\, 2$ & \multirow{2}{*}{$20 \times T \times 16$} \\
    & $3 \times 3, \text{Max pooling}, \text{stride}\, 2$ & \\

    \specialrule{.05em}{.0em}{.0em}
    \multicolumn{1}{c|}{Conv2} & $\begin{bmatrix} 3 \times 3, \, 16 \\ 
     3 \times 3, \, 16 \end{bmatrix} \times 3, \text{stride} \, 1$ & $ 20 \times T \times 16 $\\
    
    \specialrule{.05em}{.0em}{.0em}
    \multicolumn{1}{c|}{Conv3} & $\begin{bmatrix} 3 \times 3, \, 32 \\ 
     3 \times 3, \, 32 \end{bmatrix} \times 4, \text{stride} \, 2$ & $ 10 \times T/2 \times 32 $\\

    \specialrule{.05em}{.0em}{.0em}
    \multicolumn{1}{c|}{Conv4} & $\begin{bmatrix} 3 \times 3, \, 64 \\ 
     3 \times 3, \, 64 \end{bmatrix} \times 6, \text{stride} \, 2$ & $ 5 \times T/4 \times 64 $\\
    
    \specialrule{.05em}{.0em}{.0em}
    \multicolumn{1}{c|}{Conv5} & $\begin{bmatrix} 3 \times 3, 128 \\ 
     3 \times 3, 128 \end{bmatrix} \times 3, \text{stride} \, 2$ & $ 5 \times T/4 \times 128 $\\

    \specialrule{.05em}{.0em}{.0em}
    \multicolumn{1}{c|}{Pooling} & $5 \times 1 $ & $1 \times T/4 \times 128 $ \\

    \specialrule{.05em}{.0em}{.0em}
    \multicolumn{1}{c|}{Aggregation} & Self-attentive Pooling & $1 \times 128 $ \\

    \specialrule{.05em}{.0em}{.0em}
    \multicolumn{1}{c|}{FC} & Fully-connected layer, $2048$ & $1 \times 2048 $ \\

    \specialrule{.1em}{.0em}{.0em}
  \end{tabular}}
  }
  \end{table}

% %%%%%%%%%%%%%%%%%%%%%%%%%%%%%%%%%%% FIGURE 2 %%%%%%%%%%%%%%%%%%%%%%%%%%%%%%%%%%%
\begin{figure*}[t!]
\begin{minipage}[b]{1.0\linewidth}
  \centering
  \centerline{\includegraphics[width=16.2cm]{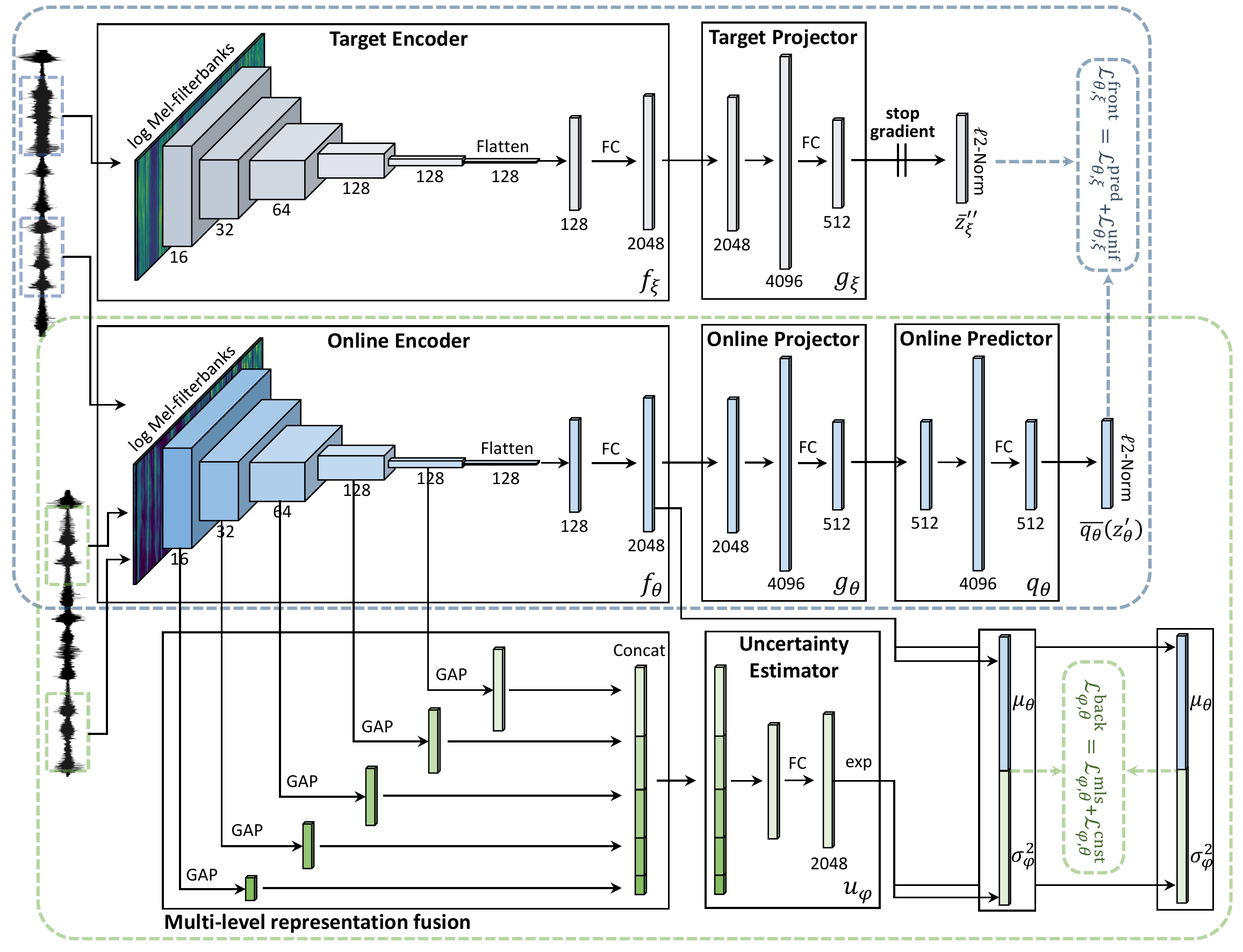}}
\vspace{0cm}
\end{minipage}
\centering
\caption{Overall process of the proposed framework. The blue block describes the architectures used in the front-end stage (the bootstrap equilibrium speaker representation learning), and the green shows the modules for the back-end stage (the uncertainty-aware probabilistic speaker embedding training).}
\label{figures}
\end{figure*}

\begin{itemize}[nosep]
\vspace{0.4mm}
\item \textbf{Online network:}
Three modules were included in the online network: the online encoder, the online projector, and the online predictor.
For the online encoder, we adopted Fast ResNet34 proposed in \cite{20Chung}.
In the speaker verification tasks, Fast ResNet34 has shown competitive performance with fewer parameters than the original ResNet of 34 layers.
We used 16, 32, 64, and 128 channels in the convolutional layers and 40-dimensional log Mel-filterbanks as the input.
The outputs were aggregated through the self-attentive pooling layer, and the 2048-dimensional speaker embeddings were extracted (shown in TABLE 1).
In the online projector, the speaker embeddings were projected to the smaller space through the multi-layer perceptron (MLP), consisting of the FC layer with 4096-dimensional output, batch normalization (BN), rectified linear units (ReLU), and the final FC layer with 512-dimensional output (i.e., FC-BN-ReLU-FC).
The 512-dimensional outputs were acquired via the final layer. For the online predictor network, the regression to predict the target projections was performed. The online predictor took the 512-dimensional online projections as the input and had the same architecture as the online projector.

\vspace{0.4mm}
\item \textbf{Target network:} In the target network, the target encoder and predictor were contained. All modules of the target network have the same architecture as the online network, except for the predictor.
The Fast ResNet34 front-end backbone was used for the target encoder, and the MLP module (i.e., FC-BN-ReLU-FC) was employed for the target projector.
The target network was composed of parameters separated from those of the online network.

\vspace{0.4mm}
\item \textbf{Uncertainty network:}
We leveraged two networks to estimate the data uncertainty: the multi-level representation fusion module and the uncertainty estimator.
The uncertainty of the speech samples can contain several factors from the low-level (e.g., frame-level local details) to high-level features (e.g., utterance-level global nuisances). To effectively process these various level features, we concatenated the five global average-pooled (GAP) convolutional layers of the online encoder in the multi-level representation fusion module.
The parameters $\theta$ of the online encoder were fixed. (shown in FIGURE 2).
For the uncertainty estimator, we utilized the MLP module with the additional exponential function layer (FC-BN-ReLU-FC-exp). The uncertainty estimator took the fused 368-dimensional multi-level representation as input and outputted the 2048-dimensional variance vector.
\end{itemize}

\subsubsection{Implementation details}
Our implementation was based on the PyTorch toolkit \cite{19Paszke} using a single NVIDIA Tesla M40 GPU with 24GB memory. During training, we randomly cropped an input utterance to two 1.80-sec segments, and then two crops were differently augmented with MUSAN noises \cite{15Snyder} and the room inverse response (RIR) filters. The MUSAN noises consisted of music, noises, and babble: 42 hours of music, 900 hours of noises, and 60 hours of babble speech.
SNR of the noises was randomly selected in the range of 0-15dB for the noise, 5-15dB for the music, and 13-20dB for the speech.
Also, for the reverberation via RIR filters, we utilized the simulated RIRs \cite{17Ko} where the filter gain was randomly sampled between -3.0 and 7.0. The 40-dimensional log Mel-filterbanks were extracted with a hamming window of 0.25-sec length and 0.10-sec hop-size with the 512-size FFT. The mean and variance normalization (MVN) was applied to the extracted acoustic features \cite{16Ulyanov}.

% %%%%%%%%%%%%%%%%%%%%%%%%%%%%%%%%%%% FIGURE 2 %%%%%%%%%%%%%%%%%%%%%%%%%%%%%%%%%%%
\begin{figure*}[t!]
     \centering
     \hfill
     \begin{subfigure}[b]{0.33\textwidth}
         \centering
         \includegraphics[width=\textwidth]{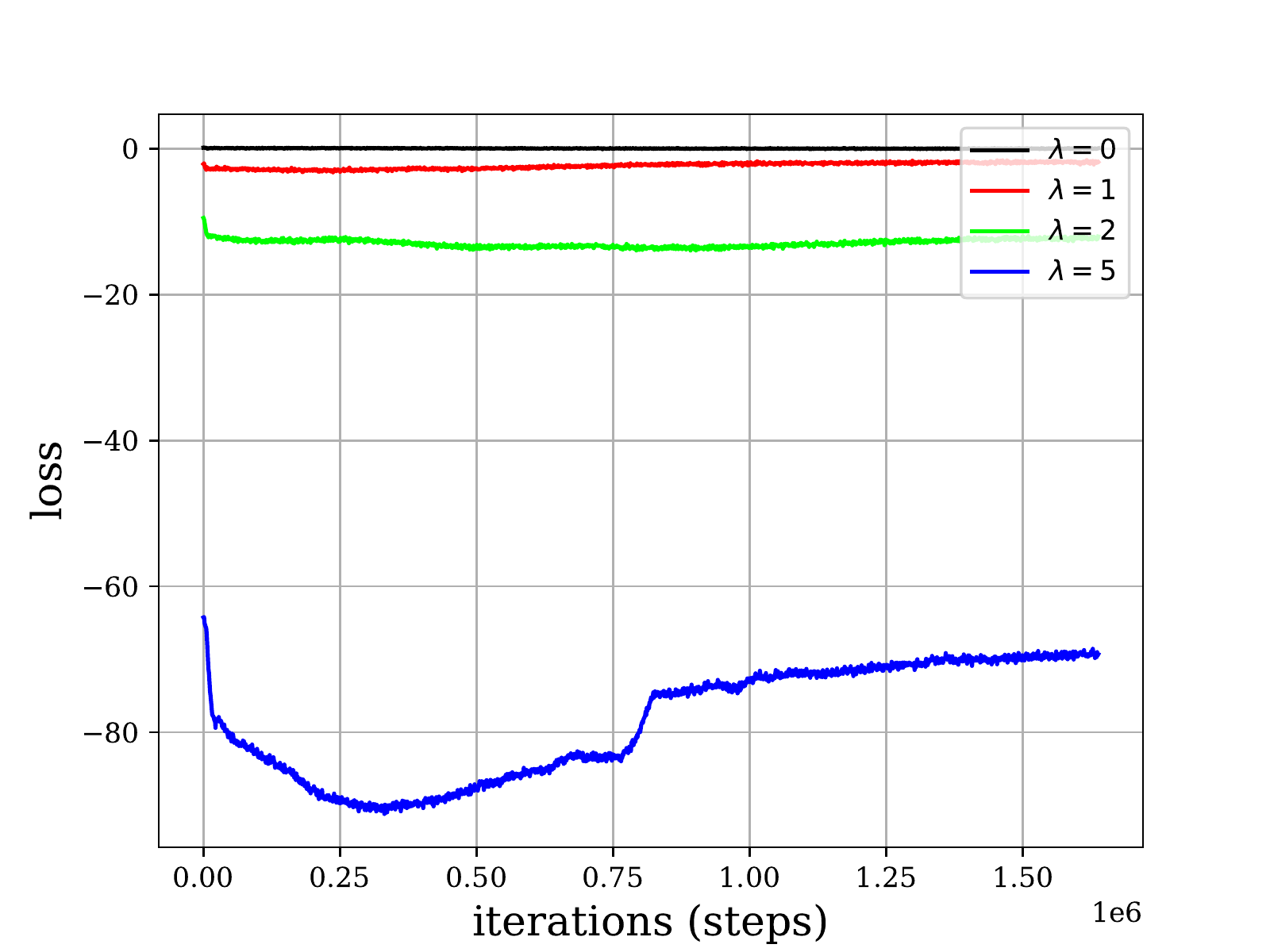}
         \caption{Front-end loss $\mathcal{L}_{\theta,\xi}^{\mathtt{front}}=\mathcal{L}_{\theta,\xi}^{\mathtt{pred}}+ \lambda \mathcal{L}_{\theta,\xi}^{\mathtt{unif}}$}
         \label{fig:}
     \end{subfigure}
     \hfill
     \begin{subfigure}[b]{0.33\textwidth}
         \centering
         \includegraphics[width=\textwidth]{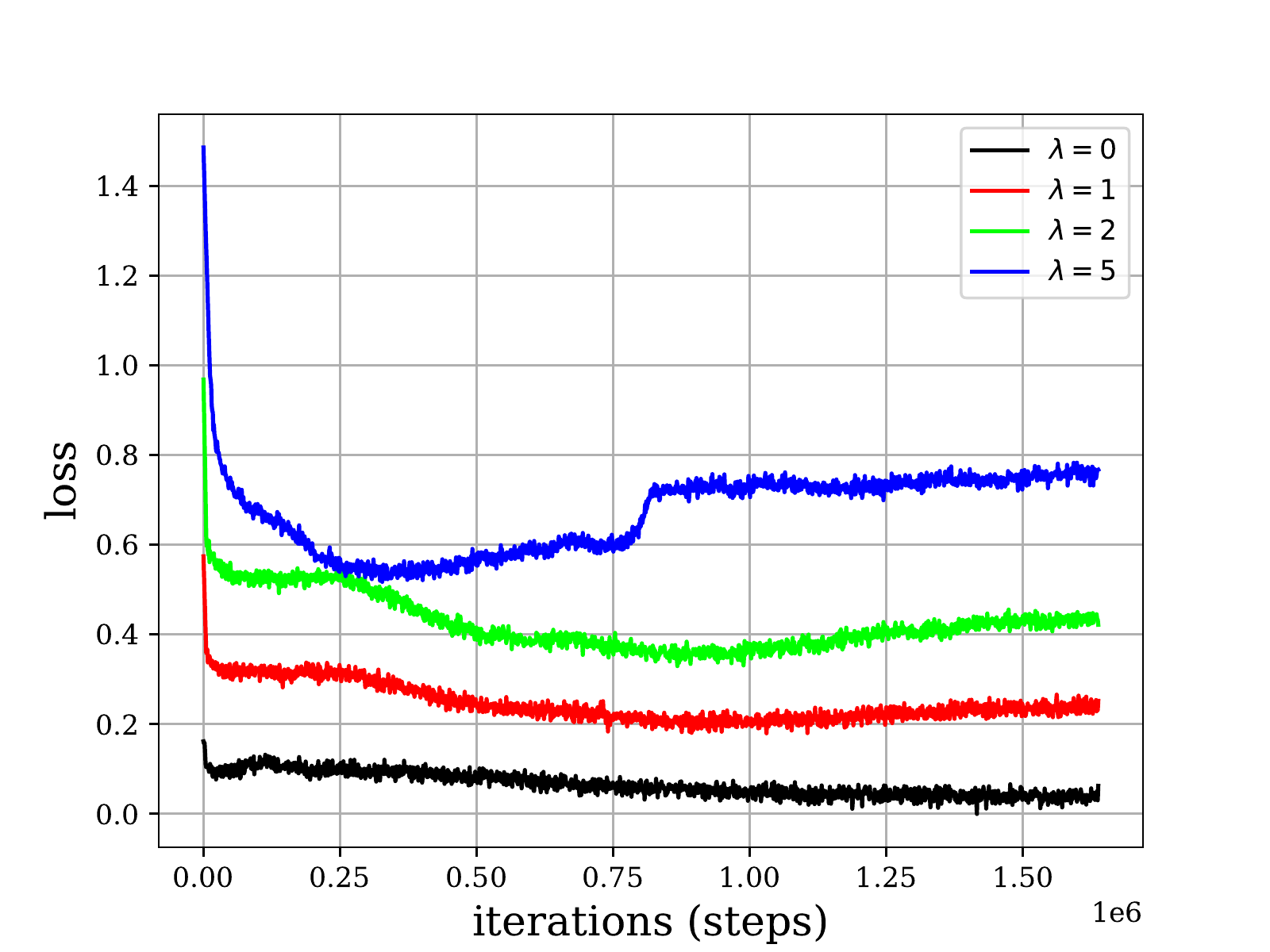}
         \caption{Bootstrap prediction loss $\mathcal{L}_{\theta,\xi}^{\mathtt{pred}}$}
         \label{fig:}
     \end{subfigure}
     \hfill
     \begin{subfigure}[b]{0.33\textwidth}
         \centering
         \includegraphics[width=\textwidth]{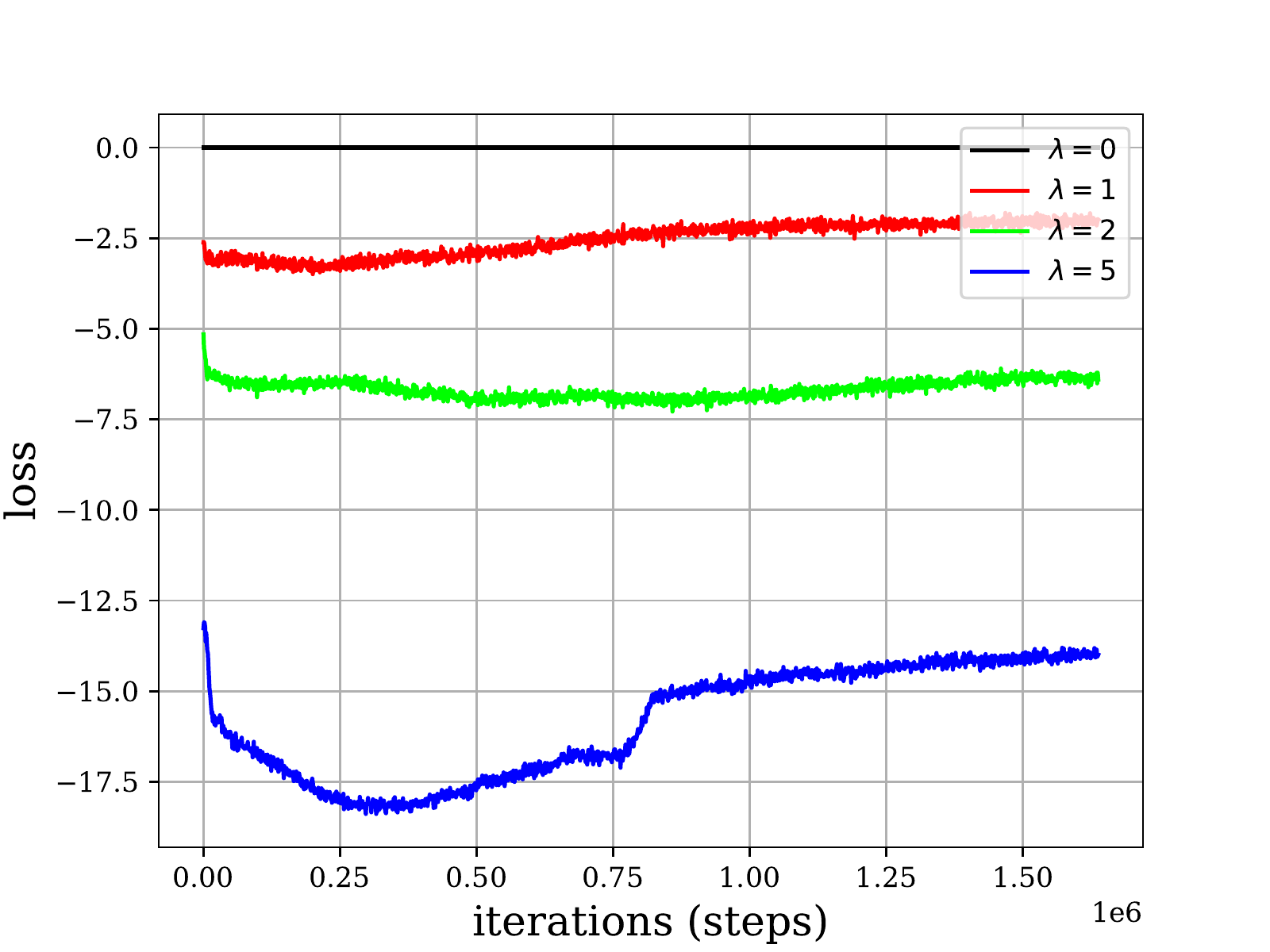}
         \caption{Uniformity regularization loss $\mathcal{L}_{\theta,\xi}^{\mathtt{unif}}$}
         \label{fig:}
     \end{subfigure}
     \hfill
     \begin{subfigure}[b]{0.33\textwidth}
         \centering
         \includegraphics[width=\textwidth]{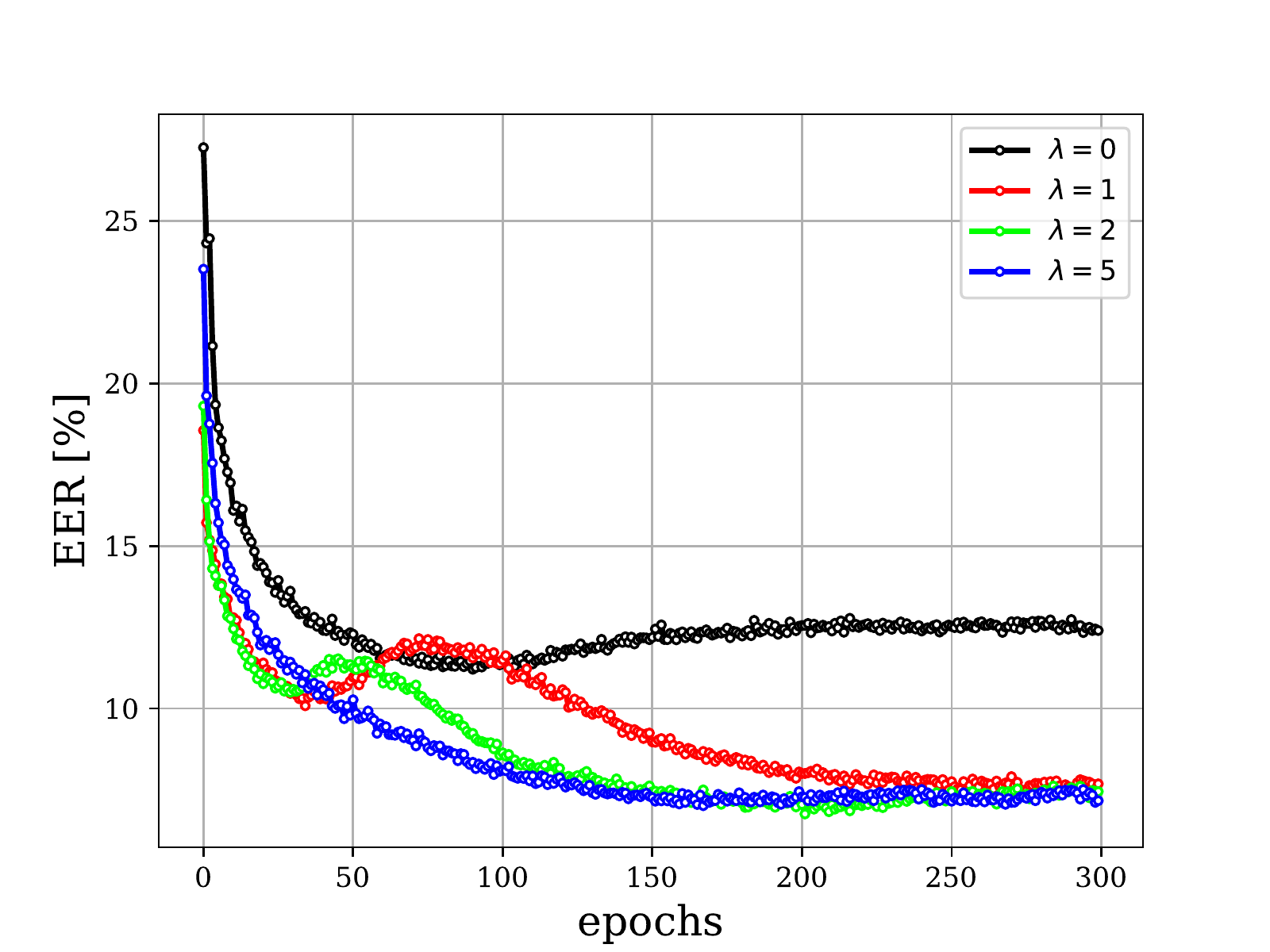}
         \caption{Equal error rate}
         \label{fig:five over x}
     \end{subfigure}
     \hfill
     \begin{subfigure}[b]{0.33\textwidth}
         \centering
         \includegraphics[width=\textwidth]{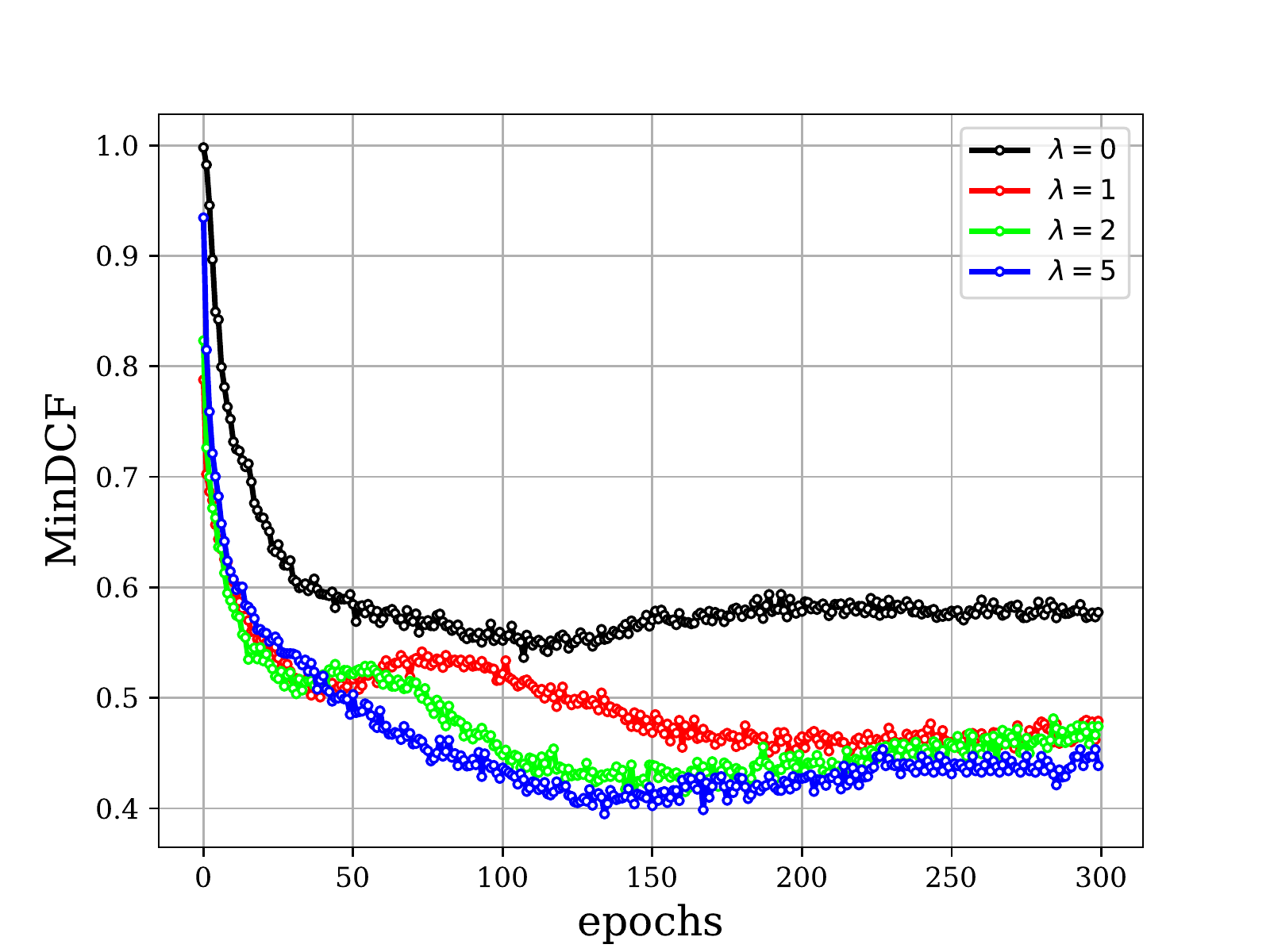}
         \caption{Minimum cost detection function}
         \label{fig:five over x}
     \end{subfigure}   
     \hfill
     \begin{subfigure}[b]{0.33\textwidth}
         \centering
         \includegraphics[width=\textwidth]{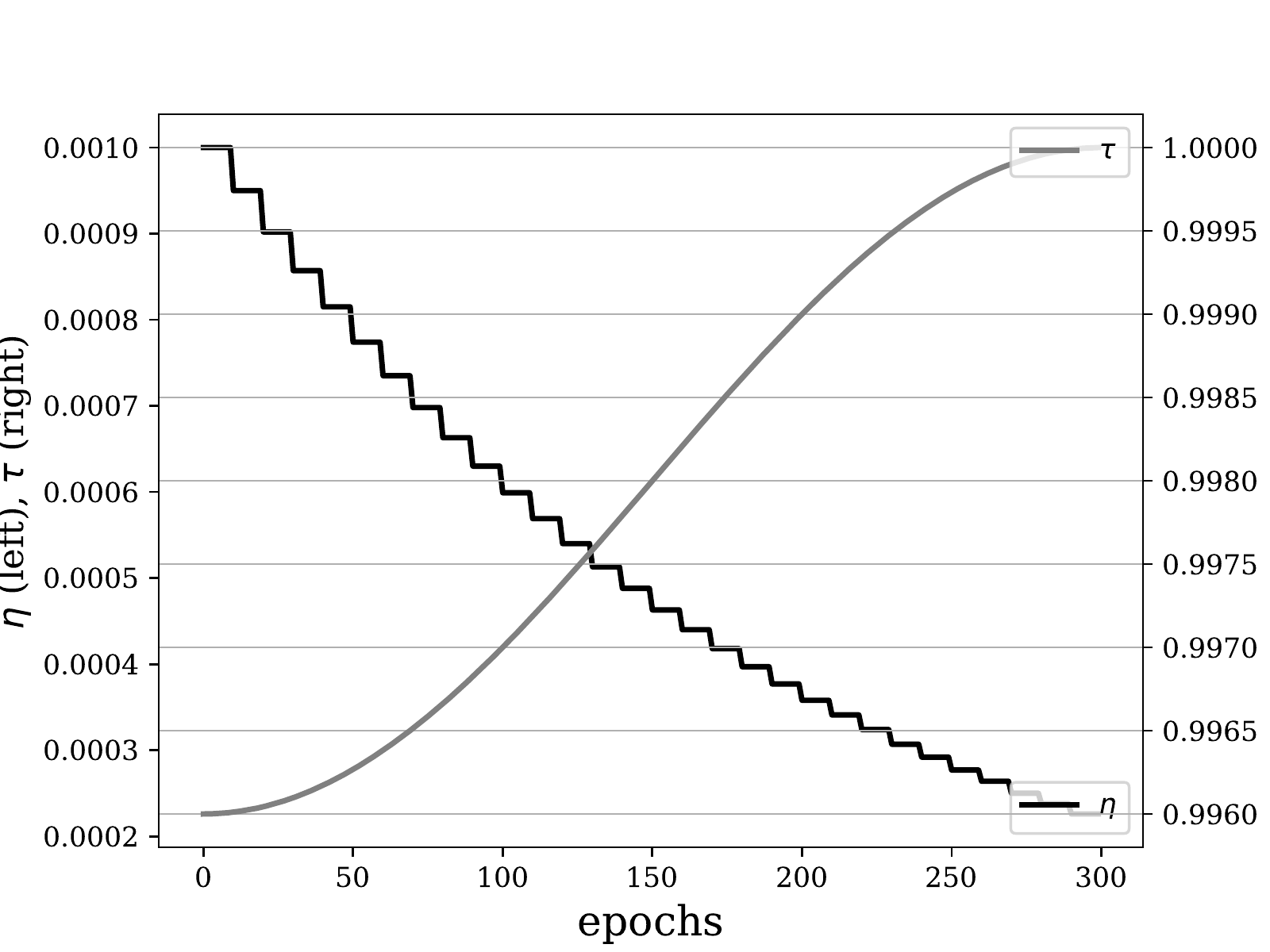}
         \caption{Learning rate $\eta$ and target decay rate $\tau$}
         \label{fig:five over x}
     \end{subfigure}
     \hfill
        \caption{(a) The front-end training loss $\mathcal{L}_{\theta, \xi}^{\mathtt{front}}$. (b) The bootstrap prediction loss $\mathcal{L}_{\theta, \xi}^{\mathtt{pred}}$. (c) The uniformity regularization loss $\mathcal{L}_{\theta, \xi}^{\mathtt{unif}}$. (d-e) The results of EER and MinDCF. (f) The values of the learning rate $\eta$ (left axis) and target decay rate $\tau$ (right axis). The graphs show the each value according to iterations (steps) or epochs with the four different $\lambda \in \{0,1,2,5\}$. All networks were trained using the VoxCeleb2, and EER \& MinDCF were evaluated on the original VoxCeleb1 test set.}
        \label{fig:three graphs}
\end{figure*}

All the experimented networks were trained with a base batch size of $200$. The online and uncertainty networks were optimized using the Adam optimizer \cite{15Kingma} with $\beta_1=0.9$, $\beta_2=0.999$, and initial learning rate $0.001$ decreasing by 5\% every 10 epochs.
The parameters $\xi$ of the target network were fixed by stopping gradient while optimizing the $\theta$ of the online network.
Then, they were updated via the momentum-based EMA with the target decay rate $\tau = 1-(1-\tau_{\mathtt{base}})\cdot(\cos(\pi k/K)+1)/2$, where $\tau_{\mathtt{base}}=0.996$ and $K$ is the maximum number of iteration steps.
For the verification, the cosine similarity and MLS were used for the score measure, and two performance metrics were evaluated: the equal error rate (EER) and the minimum detection cost function (MinDCF).
The EER indicates the error when the false alarm rate (FAR) and the false reject rate (FRR) are the same, and the MinDCF is defined as the minimum value of the weighted sum of the FAR and FRR.
The parameters of MinDCF were set as $C_{miss}=1$, $C_{fa}=1$, and $P_{target}=0.05$.

\subsection{Results: Speaker verification via bootstrap equilibrium training strategy}
\subsubsection{Training analysis}
In order to investigate whether the proposed training strategy prevents speaker representations from collapsing to a trivial solution, we conducted the ablation on the original VoxCeleb1 test set using the VoxCeleb2 training set.
Since we argued that the uniformity regularization loss $\mathcal{L}_{\theta, \xi}^{\mathtt{unif}}$ can help avoid the collapsed representations by enhancing the missing speaker-variability in the latent space, we validated the effectiveness of the uniformity regularization term by changing its weight $\lambda$ in the range of 0, 1, 2, and 5.

The training dataset was based on the development set of VoxCeleb2, and the front-end networks were trained via the bootstrap equilibrium training strategy with batch size of 200 for 300 epochs (total 1,638K steps).
The evaluation was done on the original VoxCeleb1 test set with the cosine similarity as the back-end scoring measurement.
The front-end objective function was $\mathcal{L}_{\theta, \xi}^{\mathtt{front}}=\mathcal{L}_{\theta, \xi}^{\mathtt{pred}} + \lambda \mathcal{L}_{\theta, \xi}^{\mathtt{unif}}$ where $\lambda \in \{0, 1, 2, 5\}$, and the rest of the experimental settings remain the same as Section \rom{4}-A.
FIGURE 3 shows (a) the front-end training loss $\mathcal{L}_{\theta, \xi}^{\mathtt{front}}$, (b) the bootstrap prediction loss $\mathcal{L}_{\theta, \xi}^{\mathtt{pred}}$, (c) the uniformity regularization loss $\mathcal{L}_{\theta, \xi}^{\mathtt{unif}}$, (d) the EER and (e) MinDCF results, and (f) the learning rate $\eta$ and target decay rate $\tau$ values according to the training steps or epochs.
As shown in FIGURE 3 (a) and (b), using the only bootstrap prediction loss (i.e., $\lambda=0$), $\mathcal{L}^{\mathtt{pred}}_{\theta, \xi}$ rapidly decreases to near a value of \textit{zero}; the values of this loss started from 0.1616 and reached 0.0374 at the final step (the black values in FIGURE 3 (b)).
On the other hand, jointly training with the uniformity regularization loss $\lambda \mathcal{L}^{\mathtt{unif}}_{\theta, \xi}$ prevented the bootstrap prediction loss from naively approaching a value of \textit{zero} as shown in the results of FIGURE 3 (b).
The results of EER and MinDCF in FIGURE 3 (d) and (e) showed that leveraging the proper uniformity regularization can enhance the speaker verification performance while minimizing the bootstrap prediction loss alone (i.e., $\lambda=0$) does not escape from the performance degradation.
The proposed regularization strategy can prevent representations falling into a trivial solution.
Using the proposed bootstrap equilibrium training strategy, we achieved the best performances of 6.75\% in EER with $\lambda=2$ and 0.395 in MinDCF with $\lambda=5$, which outperformed the conventional self-supervised speaker verification methods. The comparison of results with conventional methods using the VoxCeleb2 training set will be presented in TABLE 5 of Section \rom{4}-C.2.
% %%%%%%%%%%%%%%%%%%%%%%%%%%%%%%%%%%% TABLE 2 %%%%%%%%%%%%%%%%%%%%%%%%%%%%%%%%%%%
\begin{table}[t!]
\centering
\caption{Self-supervised speaker verification results on the original VoxCeleb1 test set. $\dag$: \textit{Our implementation}. $\ddag$: \textit{Proposed method}.}
\label{tab1:table}
\renewcommand{\tabcolsep}{0.7mm}
\small{
\begin{tabular}{lcccrr}
\toprule[.1em]
      \textbf{Model} & \textbf{Training} & \textbf{Augment} & \textbf{Score} & \textbf{EER} &  \textbf{DCF} \\

\midrule[.08em]
    % Supervised AProt$^{\dag}$ & VoxCeleb1 & -- & COS & 5.32 & 0.379 \\
    % \arrayrulecolor{black!40}\midrule[.04em]

    NPC \cite{19Jati} & VoxCeleb1 & -- & COS & 15.54 & 0.870 \\
    \arrayrulecolor{black!40}\midrule[.04em]         
    Prot \cite{20Lee} & VoxCeleb1 & $\mathcal{N}$ | $\mathcal{R}$ & COS & 17.42 & -- \\
    AProt \cite{20Lee} & VoxCeleb1 & $\mathcal{N}$ | $\mathcal{R}$ & COS & 14.69 & -- \\
    MoCoVox \cite{20Lee} & VoxCeleb1 & $\mathcal{N}$ | $\mathcal{R}$ & COS & 13.48 & -- \\
    
    \arrayrulecolor{black!40}\midrule[.04em]     
    AProt \cite{21Zhang} & VoxCeleb1 & $\mathcal{N}$+$\mathcal{R}$+$\mathcal{S}$ & EUC & 11.07 & 0.700 \\
    AProt + Chnl$_{\text{cos}}$ \cite{21Zhang} & VoxCeleb1 & $\mathcal{N}$+$\mathcal{R}$+$\mathcal{S}$ & EUC & 9.94 & 0.683 \\ 
    AProt + Chnl$_{\text{mse}}$ \cite{21Zhang} & VoxCeleb1 & $\mathcal{N}$+$\mathcal{R}$+$\mathcal{S}$ & EUC & 9.87 & 0.676 \\ 
    \arrayrulecolor{black!40}\midrule[.04em]     

    AProt$^{\dag}$ & VoxCeleb1& $\mathcal{N}$+$\mathcal{R}$ & COS & 11.16 & 0.696 \\    
    AProt + Unif$_{\lambda=1}$$^{\dag}$  & VoxCeleb1 & $\mathcal{N}$+$\mathcal{R}$ & COS & 10.23 & 0.549 \\ 
    \arrayrulecolor{black!40}\midrule[.04em]

    \rowcolor{mygray}
    Boot$^{\ddag}$ & VoxCeleb1 & $\mathcal{N}$+$\mathcal{R}$ & COS & 13.05 & 0.805 \\ 
    \rowcolor{mygray}
    Boot + Unif$_{\lambda=1}$$^{\ddag}$ & VoxCeleb1 &$\mathcal{N}$+$\mathcal{R}$ &COS &11.89 &0.546 \\ 
    \rowcolor{mygray}
    Boot + Unif$_{\lambda=2}$$^{\ddag}$ & VoxCeleb1 &$\mathcal{N}$+$\mathcal{R}$ &COS &9.52 &\textbf{0.485} \\ 
    \rowcolor{mygray}
    Boot + Unif$_{\lambda=5}$$^{\ddag}$ & VoxCeleb1 &$\mathcal{N}$+$\mathcal{R}$ &COS & \cellcolor{mygray}\textbf{9.20} &0.509 \\ 

    %\rowcolor{mygreen}
    \arrayrulecolor{black!40}\midrule[.04em]
    Supervised AProt$^{\dag}$ & VoxCeleb1 & -- & COS & 5.32 & 0.379 \\
    \addlinespace[.0em]
\arrayrulecolor{black}\bottomrule[.1em]
\end{tabular}
} \end{table}

\subsubsection{Comparison over the conventional methods}
In this section, we compared the performance using the VoxCeleb1 training dataset with the existing self-supervised speaker verification methods.
The previous works contain NPC \cite{19Jati}, MoCoVox \cite{20Lee}, and Channel-invariant training (Chnl) \cite{21Zhang} with prototypical (Prot) or angular prototype (AProt) loss.
NPC employed a short-term active-speaker stationarity hypothesis assuming that two temporally close speech segments belong to the same speaker and learned representations to discriminate positive and negative speaker pairs.
MoCoVox applied momentum contrast \cite{20He} on speaker representation learning using the VoxCeleb1 training set and analyzed similarity distribution for verification pairs under various augmentations.
Channel-invariant training was based on a joint training approach using the conventional angular prototypical objective and a channel-invariant loss formulated as the distance between the embedding of augmented segments and its clean version. The distance for the channel-invariant loss was computed by either the cosine similarity (Chnl$_{\text{cos}}$) or the mean squared error (Chnl$_{\text{mes}}$).
Furthermore, we included the result of the joint training for the angular prototypical and the uniformity regularization loss (AProt + Unif) with batch size of 200.
Finally, we reported the performance of the typical supervised learning method (supervised AProt \cite{20Chung}) in which AProt loss and Fast ResNet34 backbone with 512-dimensional embeddings were employed.

In this experiment, the training dataset was based on the VoxCeleb1 development set, and the data augmentation with both noise addition and reverberation ($\mathcal{N}$+$\mathcal{R}$) was used; the notation $\mathcal{N}$ | $\mathcal{R}$ denotes the data augmentation with either noise addition or reverberation, and $\mathcal{S}$ is SpecAugment \cite{19Park} in TABLE 2. The speaker representations were learned via the bootstrap equilibrium training strategy with batch size of 200 for 200 epochs (total 148.6K steps).
The evaluation was performed on the original VoxCeleb1 test set, and the cosine similarity (COS) was used as the back-end scoring metric; EUC denotes the euclidean distance in TABLE 2.
The rest of the experimental setups followed those of Section \rom{4}-A.

% %%%%%%%%%%%%%%%%%%%%%%%%%%%%%%%%%%% TABLE 3 %%%%%%%%%%%%%%%%%%%%%%%%%%%%%%%%%%%
\begin{table}[t!]
\centering
\caption{Results over different batch sizes on the original VoxCeleb1 test set. Aug: Augmentation type. BS: Batch size.}
\label{tab1:table}
\renewcommand{\tabcolsep}{0.9mm}
\small{
\begin{tabular}{c|c|c|c|r|r}
\toprule[.1em]
      \textbf{Objective} & \textbf{Training} & \textbf{Aug} & \textbf{BS} & \textbf{EER}\,\,\,\,\,\, &  \textbf{DCF}\,\,\,\,\,\, \\
\midrule[.08em]
    \multirow{4}{*}{$\mathcal{L}_{\vartheta}^{\mathtt{ap}}+\lambda\mathcal{L}_{\vartheta}^{\mathtt{unif}}$ } & VoxCeleb1 & $\mathcal{N}$+$\mathcal{R}$ & 400 & 9.88\tiny{$\pm$0.05}  & 0.519\tiny{$\pm$0.012} \\
     & VoxCeleb1 & $\mathcal{N}$+$\mathcal{R}$ & 300 & 10.01\tiny{$\pm$0.08} & 0.526\tiny{$\pm$0.007} \\
     & VoxCeleb1 & $\mathcal{N}$+$\mathcal{R}$ & 200 & 10.23\tiny{$\pm$0.04} & 0.549\tiny{$\pm$0.018} \\     
     & VoxCeleb1 & $\mathcal{N}$+$\mathcal{R}$ & 100 & 10.82\tiny{$\pm$0.10} & 0.557\tiny{$\pm$0.013} \\
    \arrayrulecolor{black}\midrule[.04em]     
  
    \multirow{4}{*}{$\mathcal{L}_{\theta, \xi}^{\mathtt{pred}}+\lambda\mathcal{L}_{\theta, \xi}^{\mathtt{unif}}$}  & VoxCeleb1 & $\mathcal{N}$+$\mathcal{R}$ & 400 & 9.05\tiny{$\pm$0.08} & 0.506\tiny{$\pm$0.008} \\
    & VoxCeleb1 & $\mathcal{N}$+$\mathcal{R}$ & 300 & 9.10\tiny{$\pm$0.05} & 0.508\tiny{$\pm$0.015} \\ 
    & VoxCeleb1 & $\mathcal{N}$+$\mathcal{R}$ & 200 & 9.20\tiny{$\pm$0.03} & 0.509\tiny{$\pm$0.009} \\    
    & VoxCeleb1 & $\mathcal{N}$+$\mathcal{R}$ & 100 & 9.38\tiny{$\pm$0.09} & 0.512\tiny{$\pm$0.014} \\ 
    
    \addlinespace[.0em]
\arrayrulecolor{black}\bottomrule[.1em]
\end{tabular}}
\end{table}
% %%%%%%%%%%%%%%%%%%%%%%%%%%%%%%%%%%% TABLE 4 %%%%%%%%%%%%%%%%%%%%%%%%%%%%%%%%%%%
\begin{table}[t!]
\centering
\caption{Relative deterioration rate on batch size decrease.}
\label{tab1:table}
\renewcommand{\tabcolsep}{0.2mm}
\small{
\begin{tabular}{c|c|c|c|c}
\specialrule{.1em}{0em}{0em}
      \textbf{Objective} & \textbf{Metric} & 400 $\rightarrow$ 300 & 300 $\rightarrow$ 200 & 200 $\rightarrow$ 100 \\
\specialrule{.08em}{0em}{0em}
    \multirow{2}{*}{$\mathcal{L}_{\vartheta}^{\mathtt{ap}}+\lambda\mathcal{L}_{\vartheta}^{\mathtt{unif}}$ } & EER drop$\downarrow$ & 1.32\% & 2.20\% & 5.77\% \\
    & DCF drop$\downarrow$ & 1.35\% & 4.37\% & 1.46\% \\
    \specialrule{.05em}{0em}{0em}
    \multirow{2}{*}{\makecell[c]{$\mathcal{L}_{\theta, \xi}^{\mathtt{pred}}+\lambda\mathcal{L}_{\theta, \xi}^{\mathtt{unif}}$} } & EER drop$\downarrow$ & \textbf{0.55\%} & \textbf{1.10\%} & \textbf{1.96\%} \\
    & DCF drop$\downarrow$ & \textbf{0.40\%} & \textbf{0.20\%} & \textbf{0.59\%} \\

\specialrule{.1em}{0em}{0em}
\end{tabular}}
\end{table}
% %%%%%%%%%%%%%%%%%%%%%%%%%%%%%%%%%%% Figure 4 %%%%%%%%%%%%%%%%%%%%%%%%%%%%%%%%%%%
\begin{figure}[t!]
     \centering
     \hfil
     \begin{subfigure}[b]{0.23\textwidth}
         \centering
         \includegraphics[width=\textwidth]{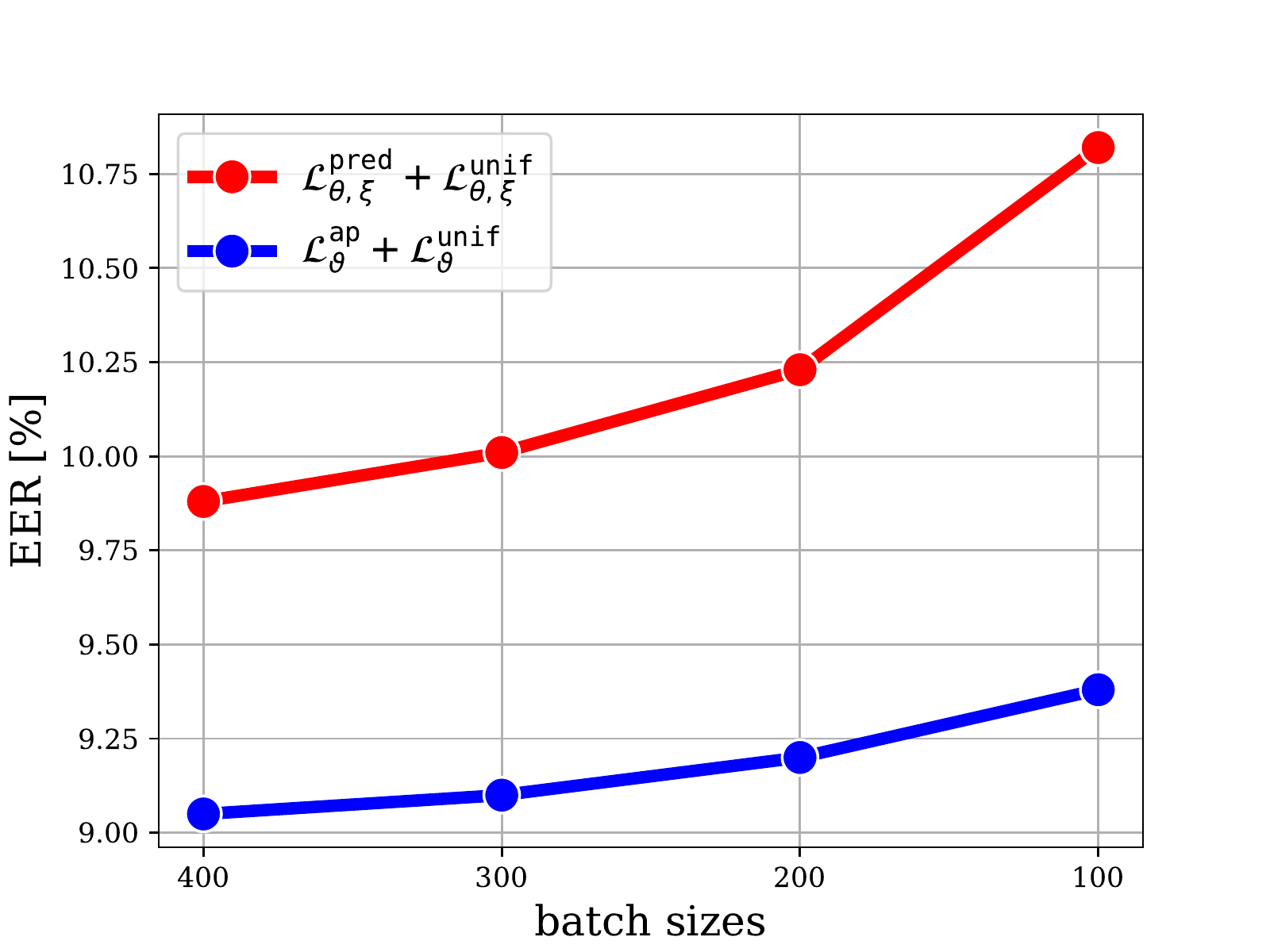}
         \caption{EER}
         \label{fig:}
     \end{subfigure}
     \hfil
     \begin{subfigure}[b]{0.23\textwidth}
         \centering
         \includegraphics[width=\textwidth]{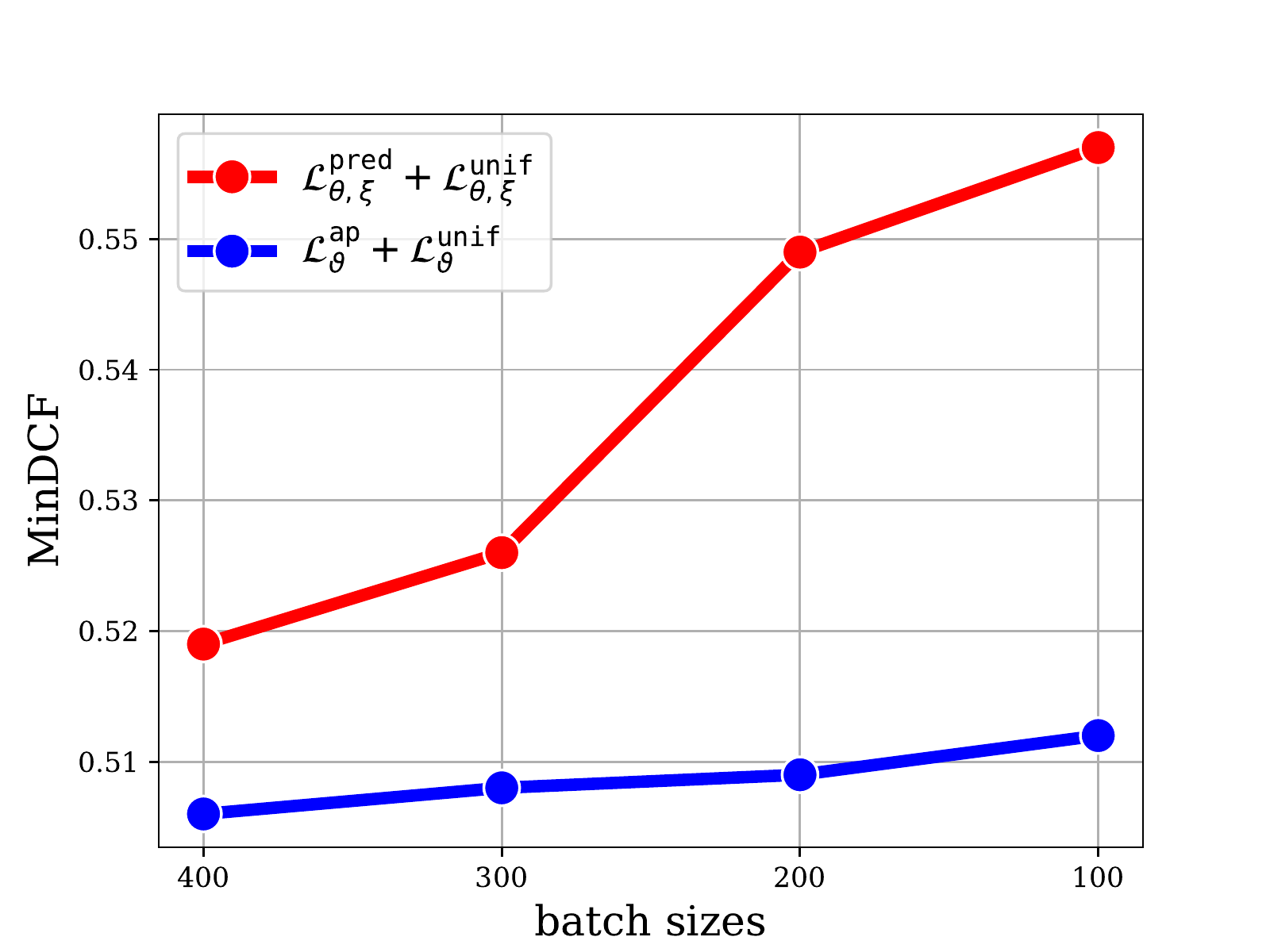}
         \caption{MinDCF}
         \label{fig:}
     \end{subfigure}
     \hfil
        \caption{EER and MinDCF over batch size changes.}
        \label{fig:three graphs}
\end{figure}

The results are given in TABLE 2. The bootstrap equilibrium speaker representations (Boot+Unif) showed the improvement of the performance compared with the conventional methods. When the uniformity regularization loss weights were set to $\lambda=5$ and $\lambda=2$, the best performances of 9.20\% in EER and 0.485 in MinDCF were achieved, respectively.
These results outperformed the conventional methods based on the contrastive learning framework.

\subsubsection{Effectiveness of batch size changes}
By learning the speaker representations through the bootstrap training framework, we could reduce the dependency on the negative pairs, which is the essential factor in contrastive learning methods. To demonstrate this effect, we compared the performance degradation as the batch size decreases with the contrastive learning method.

Two models were utilized for the comparison.
First, for the bootstrap training, we learned the speaker representations via the bootstrap equilibrium training strategy with uniformity regularization weight $\lambda=5$. The experimental setups were the same as Section \rom{4}-A, \rom{4}-B.1, and \rom{4}-B.2.

Next, to check the performance of the contrastive learning method, we used the angular prototypical loss \cite{20Huh, 21Zhang, 21Xia} and further added the uniformity regularization term. This model has shown the competitive results on both VoxCeleb1 and VoxCeleb2 (shown in Section \rom{4}-C.2) training set with relatively limited resources (e.g., augmentation types, smaller batch size, and front-end architecture with fewer parameters) compared with the previous works.

We used the VoxCeleb1 training set to train the front-end networks with a batch size of 100, 200, 300, and 400 during 148.6K iterations.
The results were evaluated on the original VoxCeleb1 test set with the cosine similarity scoring measurement.
The experiments were repeated three times, and we reported the mean and standard deviation.
The results of TABLE 3 indicate that the performance of both models gradually deteriorates as the batch size decreases. 
However, the bootstrap model showed less performance drops than the contrastive method regarding EER and MinDCF. In particular, in terms of MinDCF, consistent performances over the batch size changes were achieved. 
The relative performance deterioration rate and the summary for the batch size changes can be seen in TABLE 4 and FIGURE 4, respectively.

\subsection{Results: Uncertainty-aware probabilistic speaker embedding training strategy}
\subsubsection{Estimation of data uncertainty via the back-end network}
The uncertainty estimator is trained to capture the data uncertainty (e.g., ambient noises, reverberant environments, distortions, etc.) from the input speech samples.
To check the back-end network's ability to model the data uncertainty, we confirmed the distribution of the estimated variance output from the uncertainty estimator.
Through the trained uncertainty estimator, we inferred the uncertainty on the speech samples from the development sets of VoxCeleb1 and 2. Furthermore, we applied the five-strength random noises and reverberations to the speech samples; the ranges of 0 to 25 dB SNR for noises, 5 to 30 dB SNR for music, 13 to 28 dB SNR for babble, and -5 to 10 gains for the reverberation were divided into the five ranges in the linear scale, respectively.
FIGURE 5 shows the distribution of estimated uncertainty on VoxCeleb1 and 2 development datasets where the different colors of distribution indicate the strength of data augmentation.
It was observed that the output estimated from more strongly augmented input speech samples had greater uncertainty values and deviations.

\subsubsection{Results of the two-stage framework and comparison with the conventional methods}
This section compares the self-supervised speaker verification results on the VoxCeleb2 training dataset between the conventional methods and the proposed two-stage framework.
The conventional methods contain Disent \cite{20Nagrani2}, CDDL \cite{20Chung2}, GCL \cite{20Inoue}, I-vector \cite{11Dehak} and AAT \cite{20Huh}, Chnl \cite{21Zhang}, ProNCE \cite{21Xia} techniques with the corresponding contrastive loss (i.e., Prot, AProt, SimCLR, MoCo, and ACont).
% %%%%%%%%%%%%%%%%%%%%%%%%%%%%%%%%%%% Figure 5 %%%%%%%%%%%%%%%%%%%%%%%%%%%%%%%%%%%
\begin{figure}[t!]
     \centering
     \hfil
     \begin{subfigure}[b]{0.23\textwidth}
         \centering
         \includegraphics[width=\textwidth]{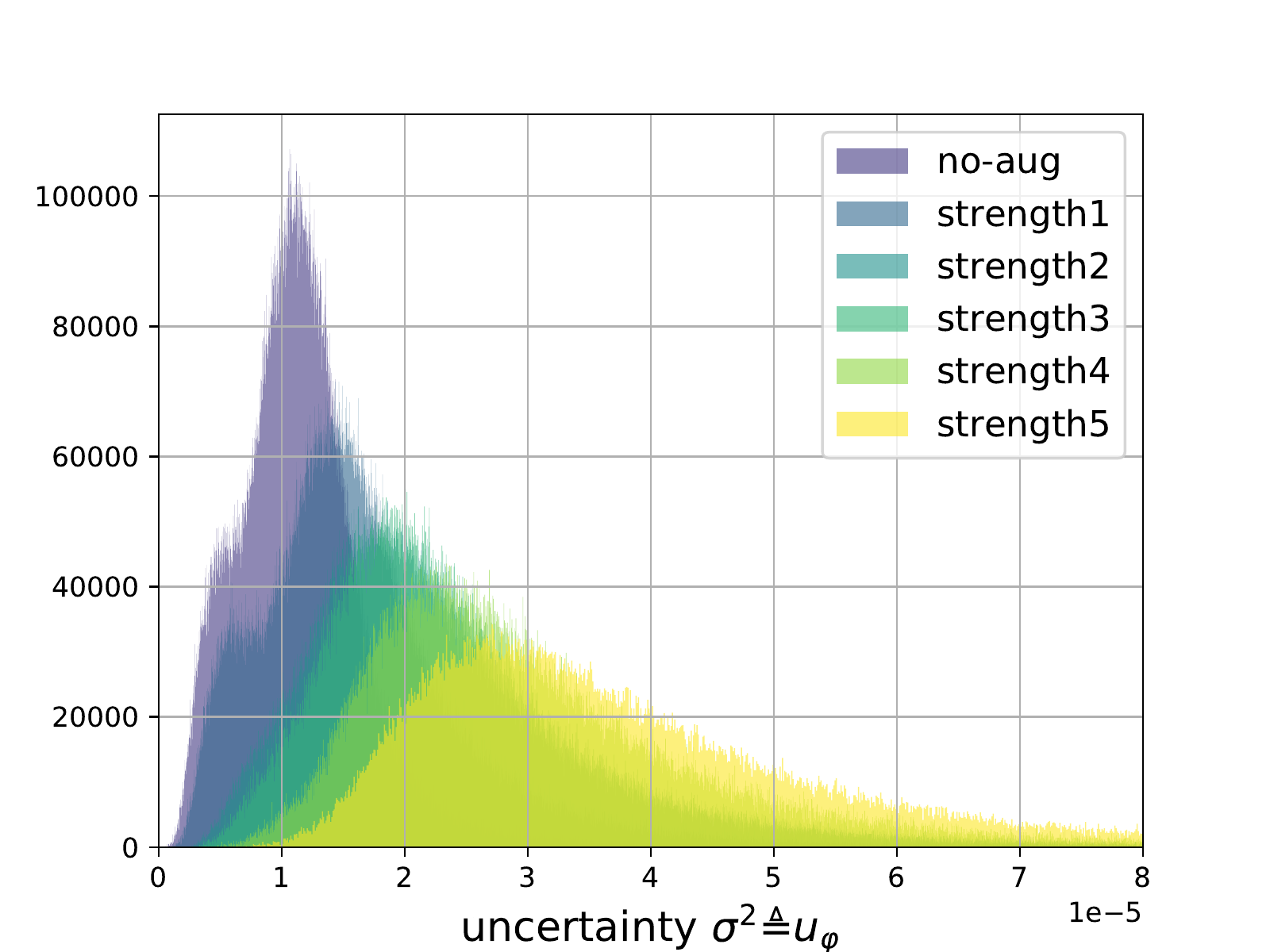}
         \caption{VoxCeleb1 dev}
         \label{fig:}
     \end{subfigure}
     \hfil
     \begin{subfigure}[b]{0.23\textwidth}
         \centering
         \includegraphics[width=\textwidth]{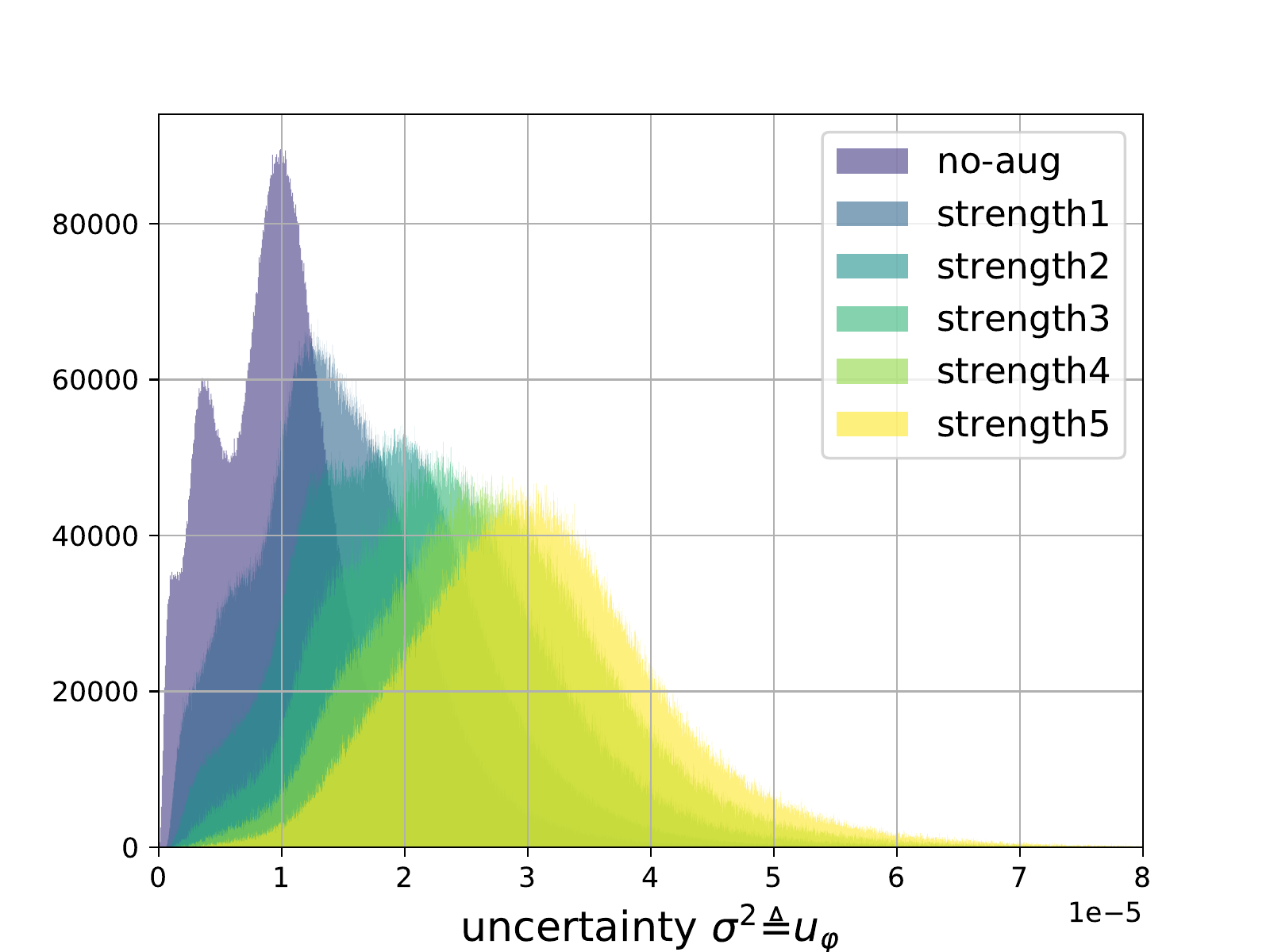}
         \caption{VoxCeleb2 dev}
         \label{fig:}
     \end{subfigure}
     \hfil
        \caption{The distribution of estimated uncertainty on the VoxCeleb1 and 2 development (dev) sets. The different colors indicate the five-strength data augmentation applied in the input speech samples.}
        \label{fig:three graphs}
\end{figure}
% %%%%%%%%%%%%%%%%%%%%%%%%%%%%%%%%%%% TABLE 5 %%%%%%%%%%%%%%%%%%%%%%%%%%%%%%%%%%%
\begin{table}[t!]
\centering
\caption{Self-supervised speaker verification results on the original VoxCeleb1 test set. MLS: Mutual likelihood score. $\dag$: \textit{Our implementation}. $\ddag$: \textit{Proposed method}.}
\label{tab1:table}
\renewcommand{\tabcolsep}{0.7mm}
\small{
\begin{tabular}{lcccrr}
\toprule[.1em]
      \textbf{Model} & \textbf{Training} & \textbf{Augment} & \textbf{Score} & \textbf{EER} &  \textbf{DCF} \\

\midrule[.08em]
    Dsient. \cite{20Nagrani2} & VoxCeleb2 & -- & COS & 22.09 & -- \\
    CDDL \cite{20Chung2} & VoxCeleb2 & -- & COS & 17.42 & -- \\
    GCL \cite{20Inoue} & VoxCeleb2 & $\mathcal{N}$ | $\mathcal{R}$ & COS & 15.26 & -- \\
    I-vector \cite{20Huh} & VoxCeleb2 & -- & COS & 15.28 & 0.627 \\
    \arrayrulecolor{black!40}\midrule[.04em]
    
    Prot \cite{20Huh} & VoxCeleb2 & $\mathcal{N}$ | $\mathcal{R}$ & EUC & 12.42 & 0.623 \\
    Prot \cite{20Huh} & VoxCeleb2 & $\mathcal{N}$+$\mathcal{R}$ & EUC & 10.16 & 0.524 \\
    AProt \cite{20Huh} & VoxCeleb2 & $\mathcal{N}$ | $\mathcal{R}$ & EUC & 11.60 & 0.620 \\
    AProt \cite{20Huh} & VoxCeleb2 & $\mathcal{N}$+$\mathcal{R}$ & EUC & 9.56 & 0.511 \\
    %\arrayrulecolor{black!40}\midrule[.04em]     

    Prot + AAT \cite{20Huh} & VoxCeleb2& $\mathcal{N}$ | $\mathcal{R}$ & EUC & 10.54 & 0.544 \\
    Prot + AAT \cite{20Huh} & VoxCeleb2& $\mathcal{N}$+$\mathcal{R}$ & EUC & 9.36 & 0.482 \\
    AProt + AAT \cite{20Huh} & VoxCeleb2& $\mathcal{N}$ | $\mathcal{R}$ & EUC & 9.03 & 0.512 \\
    AProt + AAT \cite{20Huh}  & VoxCeleb2 & $\mathcal{N}$+$\mathcal{R}$ & EUC & 8.65 & 0.454 \\
    \arrayrulecolor{black!40}\midrule[.04em]

    AProt \cite{21Zhang} & VoxCeleb2 & $\mathcal{N}$+$\mathcal{R}$+$\mathcal{S}$ & EUC & 9.23 & 0.646 \\
    AProt + Chnl$_{\text{cos}}$ \cite{21Zhang} & VoxCeleb2 & $\mathcal{N}$+$\mathcal{R}$+$\mathcal{S}$ & EUC & 8.31 & 0.615 \\ 
    AProt + Chnl$_{\text{mse}}$ \cite{21Zhang} & VoxCeleb2 & $\mathcal{N}$+$\mathcal{R}$+$\mathcal{S}$ & EUC & 8.28 & 0.610 \\ 
    \arrayrulecolor{black!40}\midrule[.04em]     

    SimCLR \cite{21Xia} & VoxCeleb2 & -- & COS & 18.14 & 0.810 \\ 
    MoCo \cite{21Xia} & VoxCeleb2 & -- & COS & 15.11 & 0.800 \\ 
    MoCo \cite{21Xia} & VoxCeleb2 & $\mathcal{S}$ & COS & 15.50 & 0.810 \\ 
    MoCo \cite{21Xia} & VoxCeleb2 & $\mathcal{N}$+$\mathcal{R}$ & COS & 8.63 & 0.640 \\ 
    MoCo (ProNCE) \cite{21Xia} & VoxCeleb2 & $\mathcal{N}$+$\mathcal{R}$ & COS & 8.23 & 0.590 \\ 
    \arrayrulecolor{black!40}\midrule[.04em]  

    ACont$^{\dag}$ & VoxCeleb2 & $\mathcal{N}$+$\mathcal{R}$ & COS & 9.49 & 0.511 \\
    AProt$^{\dag}$ & VoxCeleb2& $\mathcal{N}$+$\mathcal{R}$ & COS & 9.53 & 0.503 \\
    ACont + Unif$_{\lambda=1}$$^{\dag}$ & VoxCeleb2 & $\mathcal{N}$+$\mathcal{R}$ & COS & 8.05 & 0.455 \\
    AProt + Unif$_{\lambda=1}$$^{\dag}$ & VoxCeleb2 & $\mathcal{N}$+$\mathcal{R}$ & COS & 8.01 & 0.459 \\ 
    \arrayrulecolor{black!40}\midrule[.04em]

    \rowcolor{mygray}
    Boot$^\ddag$ & VoxCeleb2 & $\mathcal{N}$+$\mathcal{R}$ & COS & 11.20 & 0.536 \\ 
    \rowcolor{mygray}
    Boot + Unif$_{\lambda=1}$$^\ddag$ & VoxCeleb2 & $\mathcal{N}$+$\mathcal{R}$ & COS & 7.47 & 0.450 \\ 
    \rowcolor{mygray}
    Boot + Unif$_{\lambda=2}$$^\ddag$ & VoxCeleb2 & $\mathcal{N}$+$\mathcal{R}$ & COS & \textbf{6.75} & 0.414 \\ 
    \rowcolor{mygray}
    Boot + Unif$_{\lambda=5}$$^\ddag$ & VoxCeleb2 & $\mathcal{N}$+$\mathcal{R}$ & COS & 7.00 & \textbf{0.395} \\ 
    \arrayrulecolor{black!40}\midrule[.04em]

    \rowcolor{myblue}
    Boot$^\ddag$ & VoxCeleb2 & $ \mathcal{N}$+$\mathcal{R}$ & MLS & 10.83 & 0.498 \\
    \rowcolor{myblue}
    Boot + Unif$_{\lambda=1}$$^\ddag$ & VoxCeleb2 & $\mathcal{N}$+$\mathcal{R}$ & MLS & 7.08 & 0.430 \\ 
    \rowcolor{myblue}
    Boot + Unif$_{\lambda=2}$$^\ddag$ & VoxCeleb2 & $\mathcal{N}$+$\mathcal{R}$ & MLS & \textbf{6.42} & \textbf{0.345} \\ 
    \rowcolor{myblue}
    Boot + Unif$_{\lambda=5}$$^\ddag$ & VoxCeleb2 & $\mathcal{N}$+$\mathcal{R}$ & MLS & 6.71 & 0.385 \\ 

    %\rowcolor{mygreen}
    \arrayrulecolor{black!40}\midrule[.04em]
    Supervised AProt \cite{20Chung} & VoxCeleb2 & -- & COS & 2.22 & - \\

    \addlinespace[.0em]
\arrayrulecolor{black}\bottomrule[.1em]
\end{tabular}
}
\end{table}
%%%%%%%%%%%%%%%%%%%%%%%%%%%%%%%%%% TABLE 6 %%%%%%%%%%%%%%%%%%%%%%%%%%%%%%%%%%%
\begin{table*}[t!]
\centering
\caption{The self-supervised speaker verification performance of the two-stage framework. The VoxCeleb1\&2 sets were used for the front- and back-end networks training, and the original VoxCeleb1 test set was employed for the evaluation.}
\label{tab3:table}
\renewcommand{\tabcolsep}{1.4mm}
\small{
\begin{tabular}{c|ccc||c|cccc||cc} %L{2.85cm} C{0.5cm}
\toprule[.1em]
    \multicolumn{4}{c||}{\multirow{1}{*}{\textbf{Front-end}}}
    & \multicolumn{5}{c||}{\multirow{1}{*}{\textbf{Back-end}}} 
    & \multicolumn{2}{c}{\textbf{Evaluation}} \\
        \cmidrule(lr){1-4} \cmidrule(lr){5-9} \cmidrule(lr){10-11}
        \textbf{Objective}
        & \textbf{Training}
        & \textbf{Augment}
        & \textbf{Unif.}
        & \textbf{Objective}
        & \textbf{Training}
        & \textbf{Augment}
        & \textbf{Cnst.}
        & \textbf{Score}
        & \textbf{EER}
        & \textbf{DCF} \\
\midrule[.08em]

    \multirow{12}{*}{$\mathcal{L}_{\theta, \xi}^{\mathtt{pred}}+\lambda\mathcal{L}_{\theta, \xi}^{\mathtt{unif}}$} % 24 -> 12
    & VoxCeleb1 & $\mathcal{N}$+$\mathcal{R}$ & $\lambda=5.0$ & -- & -- & -- & -- & COS & 9.20 & 0.509 \\
    %-----------------------------------------------
    \arrayrulecolor{black!40}\cmidrule[.04em]{2-11}
    & VoxCeleb1 & $\mathcal{N}$+$\mathcal{R}$ & $\lambda=5.0$ & \multirow{5}{*}{$\mathcal{L}_{\varphi, \theta}^{\mathtt{mls}} + \gamma \mathcal{L}_{\varphi, \theta}^{\mathtt{cnst}}$} % 10 -> 5
                                                                 & VoxCeleb1 & $\mathcal{N}$+$\mathcal{R}$ & $\gamma=0.0$ & MLS   & 8.89 & 0.504 \\
    & VoxCeleb1 & $\mathcal{N}$+$\mathcal{R}$ & $\lambda=5.0$ &    & VoxCeleb1 & $\mathcal{N}$+$\mathcal{R}$ & $\gamma=0.5$ & MLS   & \textbf{8.84} & 0.508 \\
    & VoxCeleb1 & $\mathcal{N}$+$\mathcal{R}$ & $\lambda=5.0$ &    & VoxCeleb1 & $\mathcal{N}$+$\mathcal{R}$ & $\gamma=1.0$ & MLS   & 8.89 & 0.495 \\
    & VoxCeleb1 & $\mathcal{N}$+$\mathcal{R}$ & $\lambda=5.0$ &    & VoxCeleb1 & $\mathcal{N}$+$\mathcal{R}$ & $\gamma=2.0$ & MLS   & 8.90 & \textbf{0.489} \\
    & VoxCeleb1 & $\mathcal{N}$+$\mathcal{R}$ & $\lambda=5.0$ &    & VoxCeleb1 & $\mathcal{N}$+$\mathcal{R}$ & $\gamma=3.0$ & MLS   & 8.87 & 0.492 \\
    \arrayrulecolor{black}\cmidrule[.08em]{2-11}

    & VoxCeleb2 & $\mathcal{N}$+$\mathcal{R}$ & $\lambda=2.0$ & -- & -- & -- & -- & COS & 6.75 & 0.414 \\
    \arrayrulecolor{black!40}\cmidrule[.04em]{2-11}
    & VoxCeleb2 & $\mathcal{N}$+$\mathcal{R}$ & $\lambda=2.0$ & \multirow{5}{*}{$\mathcal{L}_{\varphi, \theta}^{\mathtt{mls}} + \gamma \mathcal{L}_{\varphi, \theta}^{\mathtt{cnst}}$} % 10 -> 5
                                                                 & VoxCeleb2 & $\mathcal{N}$+$\mathcal{R}$ & $\gamma=0.0$ & MLS   & 6.45   & 0.356 \\
    & VoxCeleb2 & $\mathcal{N}$+$\mathcal{R}$ & $\lambda=2.0$ &    & VoxCeleb2 & $\mathcal{N}$+$\mathcal{R}$ & $\gamma=0.5$ & MLS   & \textbf{6.38}   & 0.354 \\
    & VoxCeleb2 & $\mathcal{N}$+$\mathcal{R}$ & $\lambda=2.0$ &    & VoxCeleb2 & $\mathcal{N}$+$\mathcal{R}$ & $\gamma=1.0$ & MLS   & 6.42   & \textbf{0.345} \\
    & VoxCeleb2 & $\mathcal{N}$+$\mathcal{R}$ & $\lambda=2.0$ &    & VoxCeleb2 & $\mathcal{N}$+$\mathcal{R}$ & $\gamma=2.0$ & MLS   & 6.43   & 0.349 \\
    & VoxCeleb2 & $\mathcal{N}$+$\mathcal{R}$ & $\lambda=2.0$ &    & VoxCeleb2 & $\mathcal{N}$+$\mathcal{R}$ & $\gamma=3.0$ & MLS   & 6.47   & 0.358 \\
%-----------------------------------------------
\arrayrulecolor{black}\bottomrule[.1em]
\end{tabular}}
\end{table*}
Disent. and CDDL leveraged the cross-modal synchrony between faces and audio in a video for learning the speaker representations.
I-vector was a popular method in speaker recognition before the emergence of deep learning and was commonly used with probabilistic linear discriminant analysis (PLDA) back-end \cite{06Ioffe}.
In this benchmark reported in \cite{20Huh}, the cosine similarity back-end was used for the unsupervised setting.
GCL, AAT, Chnl, and ProNCE exploited data augmentation to build the self-supervision pretext task.
AAT and Chnl learned the channel-invariant speaker embeddings by using adversarial training and joint training, respectively.
ProNCE introduced a prototypical memory bank to the speaker embedding training for treating the negative samples efficiently.
Also, we included the joint training methods using the contrastive losses (AProt and ACont) and the uniformity regularization loss (Unif), where ACont loss is a symmetrical form of AProt in terms of a query and support.
These models were trained with batch size of 200 for 300 epochs using the Fast ResNet34.
Finally, we contained the performance of the fully supervised method (supervised AProt) reported in \cite{20Chung}.

In our front-end stage, the speaker representations were learned via bootstrap equilibrium training with batch size of 200 for 300 epochs, same as Section \rom{4}-B.2.
Moreover, in the back-end training stage, we fixed their parameters and utilized them as mean vectors of the probabilistic speaker embeddings. Finally, the uncertainty estimator was trained with batch size of 200 for 30 epochs (total 163.8K steps). All networks were trained using the VoxCeleb2 training set, and the objective was set as equation (23), i.e., $\mathcal{L}_{\varphi, \theta}^{\mathtt{back}} = \mathcal{L}_{\varphi, \theta}^{\mathtt{mls}} + \gamma \mathcal{L}_{\varphi, \theta}^{\mathtt{cnst}}$, where $\gamma$ was set to $1$.
The results were evaluated on the original VoxCeleb1 test set, and the mutual likelihood score (MLS) was used as the measurement with the mean and variance of probabilistic speaker embeddings learned in the back-end stage.

The results are shown in TABLE 5. First, the performance of the bootstrap equilibrium speaker representations with cosine similarity back-end outperformed the conventional self-supervised speaker representation methods.
The best performing model achieved the 6.75\% in EER and 0.395 in MinDCF with $\lambda=2$ and $5$, respectively. Next, by performing the evaluation through MLS between the estimated probabilistic speaker embeddings, the performance was further improved.
The best performance showed the 6.42\% in EER and 0.345 in MinDCF with $\lambda=2$.

\subsubsection{Performance analysis of the back-end stage}
In order to investigate the speaker verification performance in the back-end stage, we conducted ablation on the original VoxCeleb1 test set. 
First, we checked the speaker verification results on the VoxCeleb1 and VoxCeleb2 training datasets, respectively. The best performing models of the bootstrap equilibrium speaker embeddings in the front-end, i.e., $\lambda=5$ for the VoxCeleb1 and $\lambda=2$ for the VoxCeleb2, were used as the mean vector of the probabilistic speaker embeddings. Also, we analyzed the performance on the different uncertainty constraint loss weights $\gamma$ in the range of 0, 0.5, 1, 2, and 3.
In the front-end stage, the online encoder was trained with batch size of 200 for 200 epochs on the VoxCeleb1 dataset. In the VoxCeleb2, the network was trained for 300 epochs. Also, the uncertainty network was trained with the same batch size for 30 epochs in both training sets.
As shown in TABLE 6, The speaker verification performances on both VoxCeleb1 and 2 training sets were further enhanced, which showed the best results of 8.84\% in EER and 0.489 in MinDCF on the VoxCeleb1, and 6.38\% in EER and 0.345 in MinDCF on the VoxCeleb2.
Compared to a corresponding cosine similarity back-end performance, we could achieve the relative improvements of 3.91\% in EER and 3.93\% in MinDCF on the VoxCeleb1, and 5.48\% in EER and 16.67\% in MinDCF on the VoxCeleb2, respectively.

\section{Conclusion}
In this paper, we proposed self-supervised speaker representation learning strategies, consisting of the \textit{bootstrap equilibrium speaker representation learning} in the front-end and the \textit{uncertainty-aware probabilistic speaker embedding training} in the back-end.
For the front-end stage, we learned the speaker representations via the bootstrap training scheme with the uniformity regularization term. Then, in the back-end stage, the probabilistic speaker embedding was estimated by maximizing the MLS.
Finally, we computed the MLS between the estimated probabilistic speaker embeddings and utilized them for the verification.
The integrated two-stage framework showed outstanding results, outperforming the conventional methods based on contrastive learning.

\begin{IEEEbiography}[{\includegraphics[width=1in,height=1.25in,clip,keepaspectratio]{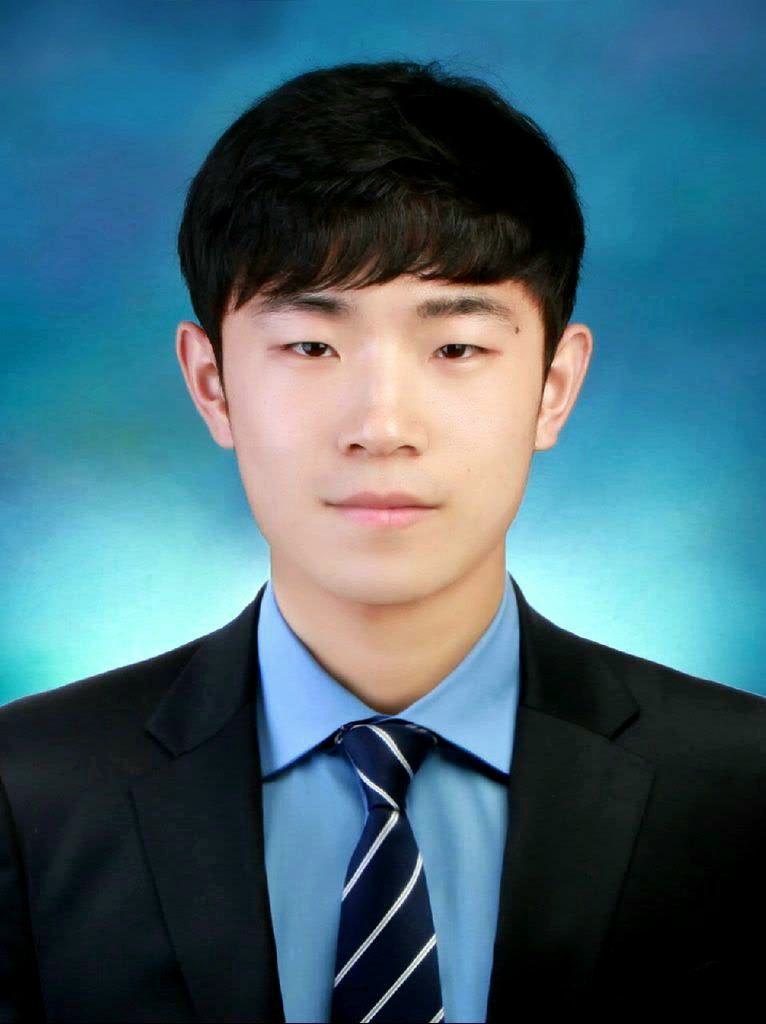}}]{SUNG HWAN MUN} was born in Incheon, Korea, in 1993. He received the B.S. degree in electronics engineering from Inha University, Incheon, Korea, in 2017. He is currently pursuing the Ph.D. degree in electrical and computer engineering at Seoul National University (SNU), Seoul, Korea.

His research interests include speaker recognition, machine learning, and signal processing.
\end{IEEEbiography}

\begin{IEEEbiography}[{\includegraphics[width=1in,height=1.25in,clip,keepaspectratio]{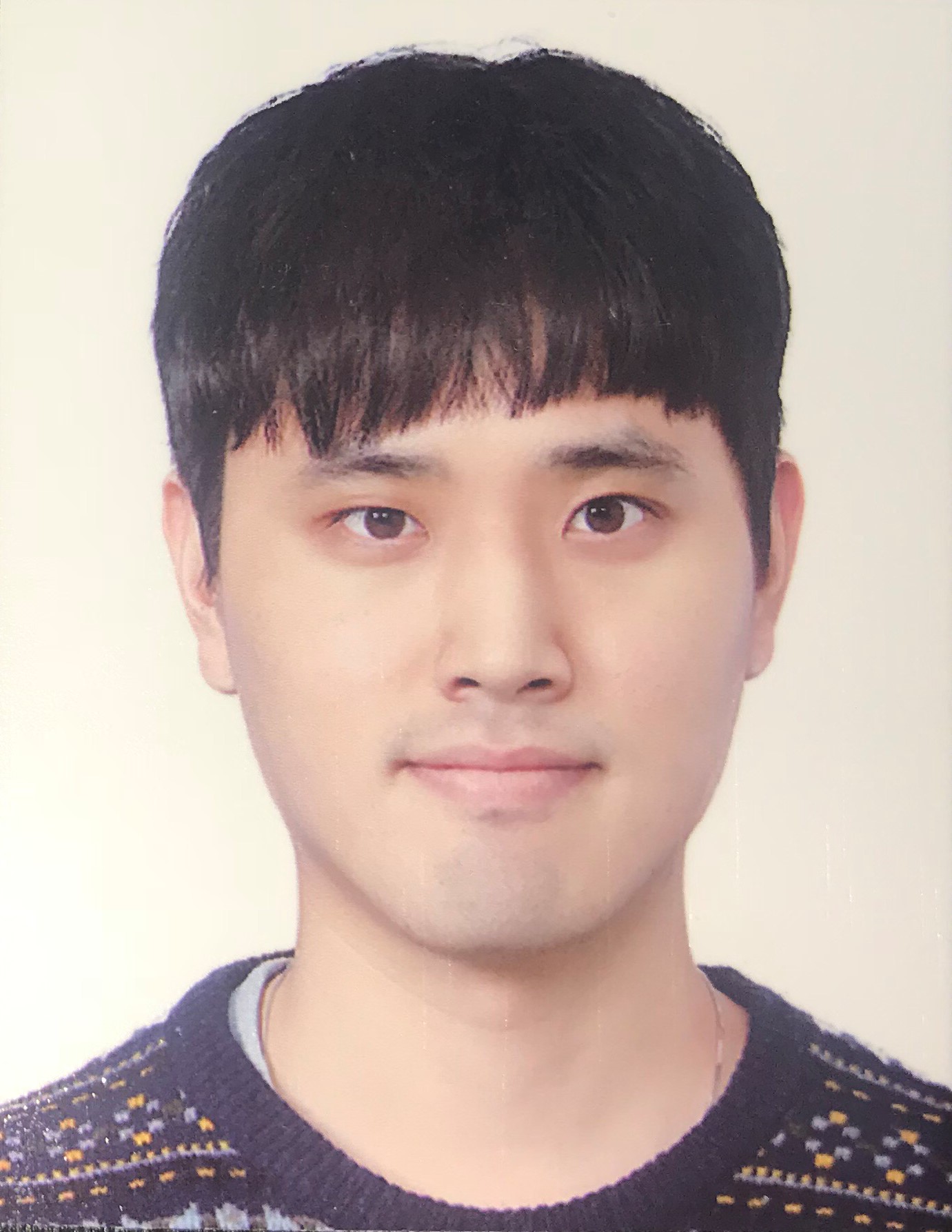}}]{MIN HYUN HAN} was born in Seoul, Korea, in 1992. He received the B.S. degree in electrical \& electronic engineering from Yonsei University, Seoul, Korea, in 2018. He is currently pursuing the Ph.D. degree in electrical and computer engineering at Seoul National University (SNU), Seoul, Korea.

His research interests include speaker recognition, machine learning, and signal processing.
\end{IEEEbiography}

\begin{IEEEbiography}[{\includegraphics[width=1in,height=1.25in,clip,keepaspectratio]{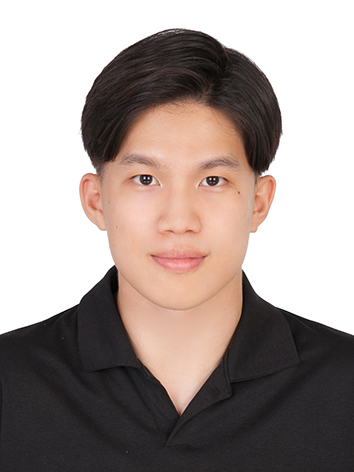}}]{DONGJUNE LEE} received the B.S. degree in Electrical Engineering from Korea University, Seoul, Korea in 2020. He is currently pursuing the Ph.D. degree in electrical and computer engineering at Seoul National University (SNU), Seoul, Korea.

His research interests include speech recognition, anomaly detection, and machine learning.
\end{IEEEbiography}

\begin{IEEEbiography}[{\includegraphics[width=1in,height=1.25in,clip,keepaspectratio]{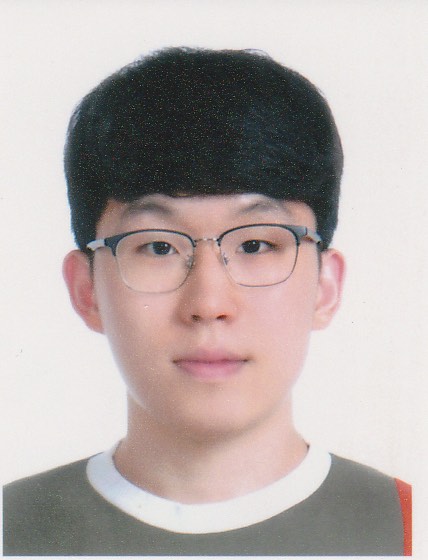}}]{JIHWAN KIM} was born in Jeonju, Korea, in 1997. He received the B.S. degree in electrical and computer engineering and the B.A. degree in linguistics from Seoul National University (SNU), Seoul, Korea, in 2021. He is currently pursuing the M.S. degree in electrical and computer engineering at Seoul National University (SNU), Seoul, Korea.

His research interests include speech enhancement, natural language processing and machine learning.
\end{IEEEbiography}

\begin{IEEEbiography}[{\includegraphics[width=1in,height=1.25in,clip,keepaspectratio]{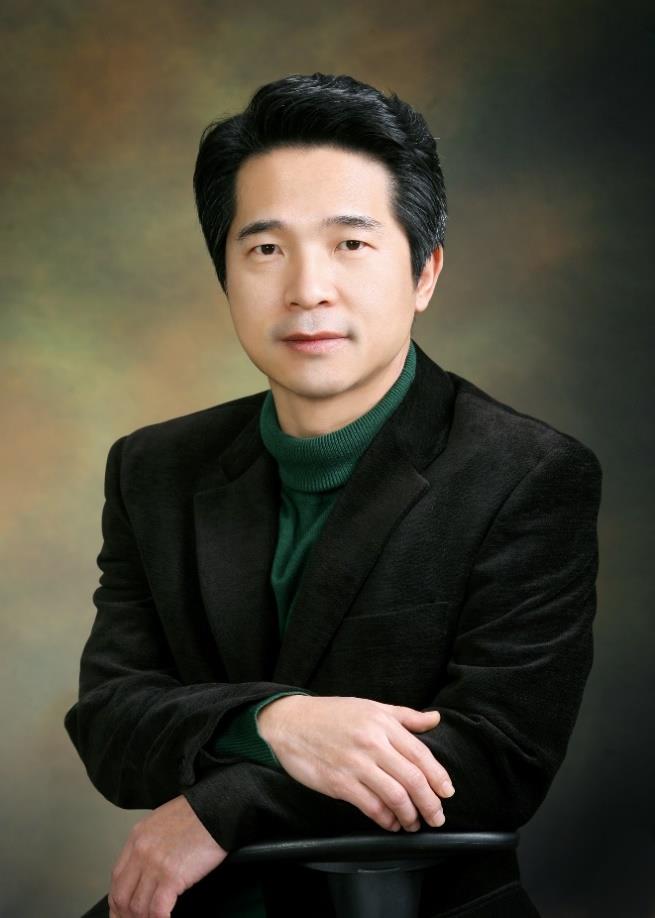}}]{NAM SOO KIM} received the B.S. degree in electronics engineering from Seoul National University (SNU), Seoul, Korea, in 1988 and the M.S. and Ph.D. degrees in electrical engineering from Korea Advanced Institute of Science and Technology in 1990 and 1994, respectively.

From 1994 to 1998, he was with Samsung Advanced Institute of Technology as a Senior Member of Technical Staff. Since 1998, he has been with the School of Electrical Engineering, SNU, where he is currently a Professor. His research area includes speech signal processing, speech recognition, speech/audio coding, speech synthesis,
adaptive signal processing, machine learning, and mobile communication.

\end{IEEEbiography}

\EOD


\begin{thebibliography}{00}
%1
\bibitem{15Hansen} J. Hansen and T. Hasan, ``Speaker recognition by machines and humans,'' \textit{IEEE Signal Processing Magazine}, vol. 32, no. 6, pp. 74–99, Oct. 2015.
%2
\bibitem{18Snyder} D. Snyder et al., ``X-vectors: Robust dnn embeddings for speaker recognition,'' in \textit{Proc. ICASSP}, IEEE, 2018.
%3
\bibitem{18Wan} L. Wan et al., ``Generalized end-to-end loss for speaker verification,'' in \textit{Proc. ICASSP}, IEEE, 2018, pp. 4879–4883.
%4
\bibitem{18Okabe} K. Okabe et al., ``Attentive statistics pooling for deep speaker embedding,'' in \textit{Proc. INTERSPEECH}, 2018, pp. 3573–3577.
%5
\bibitem{20Chung} J. S. Chung, J. Hur, S. Mun, M. Lee, H. S. Heo, S. Choe, C. Ham, S. Jung, B.-J. Lee, and I. Han, ``In defence of metric learning for speaker recognition,'' in \textit{Proc. INTERSPEECH}, 2020.
%6
\bibitem{19Jung} J. Jung, H. Heo, J. Kim, H. Shim, and H. Yu, ``RawNet: Advanced end-to-end deep neural network using raw waveforms for text-independent speaker verification,'' in \textit{Proc. INTERSPEECH}, 2019.
%7
\bibitem{17Nagrani} A. Nagrani et al., ``VoxCeleb: A large-scale speaker identification dataset,'' in \textit{Proc. INTERSPEECH}, 2017, pp. 2616–2620.
%8
\bibitem{18Chung} J. S. Chung et al., ``VoxCeleb2: Deep speaker recognition,'' in \textit{Proc. INTERSPEECH}, 2018, pp. 1086–1090.
%9
\bibitem{20Nagrani} A. Nagrani et al. ``Voxceleb: Large-scale speaker verification in the wild,'' \textit{Computer Speech \& Language} vol 60, pp. Mar. 2020.
%10
\bibitem{21Le-Khac} P. H. Le-Khac et al., ``Contrastive representation learning: A framework and review," \textit{IEEE Access}, pp. 6723-6727, 2021. vol. 19, no. 4, pp. 788--798, 2010.
%11
\bibitem{21Liu} X. Liu et al, ``Self-supervised learning: Generative or contrastive," \textit{IEEE Transactions on Knowledge and Data Engineering}, no. 01, 2021.
%12
\bibitem{18Oord} A. v. d. Oord, Y. Li, and O. Vinyals, ``Representation learning with contrastive predictive coding,'' \textit{arXiv preprint arXiv:1807.03748}, 2018.
%13
\bibitem{20Chen} T. Chen et al., ``A simple framework for contrastive learning of visual representations,'' in \textit{Proc. ICML}, PMLR, 2020, pp. 1597-1607.
%14
\bibitem{20He} K. He et al., ``Momentum contrast for unsupervised visual representation learning,'' in \textit{Proc. CVPR}, IEEE, 2020, pp. 9729-9738.
%15
\bibitem{19Arora} S. Arora et al., “A theoretical analysis of contrastive unsupervised representation learning,” \textit{arXiv preprint arXiv:1902.09229}, 2019.
%16
\bibitem{19Stafylakis} T. Stafylakis et al., ``Self-supervised speaker embeddings,'' in \textit{Proc. INTERSPEECH}, 2019.
%17
\bibitem{19Ravanelli} M. Ravanelli and Y. Bengio, ``Learning speaker representations with mutual information,'' in \textit{Proc. INTERSPEECH}, 2019, pp. 1153–1157.
%18
\bibitem{19Jati} A. Jati and P. Georgiou, ``Neural predictive coding using convolutional neural networks toward unsupervised learning of speaker characteristics,'' \textit{IEEE/ACM Transactions on Audio, Speech, and Language Processing}, vol. 27, no. 10, pp. 1577–1589, 2019.
%19
\bibitem{20Inoue} N. Inoue and K. Goto, ``Semi-supervised contrastive learning with generalized contrastive loss and its application to speaker recognition,'' in \textit{Proc. APSIPA}, IEEE, 2020.
%20
\bibitem{20Huh} J. Huh et al., ``Augmentation adversarial training for unsupervised speaker recognition,'' in \textit{Workshop on Self-upervised Learning for Speech and Audio Processing, NeurIPS}, 2020.
%21
\bibitem{21Zhang} H. Zhang et al., ``Contrastive self-supervised learning for text-independent speaker verification," in \textit{Proc. ICASSP}, IEEE, 2021, pp. 6713-6717.
%22
\bibitem{21Xia} W. Xia et al., ``Self-supervised text-independent speaker verification using prototypical momentum contrastive learning," in \textit{Proc. ICASSP}, IEEE, 2021, pp. 6723-6727.
%23
\bibitem{20Tian} Y. Tian et al., ``What makes for good views for contrastive learning,`` in \textit{Proc. NeurIPS}, 2020.
%24
\bibitem{21Robinson} J. Robinson et al., ``Contrastive learning with hard negative samples," in \textit{Proc. ICLR}, 2021.
%25
\bibitem{20Guo} Z. D. Guo et al., ``Bootstrap latent-predictive representations for multitask reinforcement learning," in \textit{Proc. ICML}, PMLR, 2020, pp. 3875-3886.
%26
\bibitem{20Grill} J.-B. Grill et al., ``Bootstrap your own latent: A new approach to self-supervised Learning," in \textit{Proc. NeuIPS}, 2020. 3875-3886.
%27
\bibitem{20Chen2} X. Chen et al., ``Exploring simple siamese representation learning,'' in \textit{Proc. CVPR}, IEEE, 2021, 15750-15758.
%28
\bibitem{21Thakoor} S. Thakoor et al., ``Bootstrapped representation learning on graphs,'' in \textit{Workshop on Geometrical and Topological Representation Learning, ICLR}, 2021.
%29
\bibitem{21Che} F. Che et al. ``Self-supervised graph representation learning via bootstrapping,'' \textit{Neurocomputing}, vol. 456, pp. 88-96, 2021.
%30
\bibitem{21Zhang2} Y. Zhang et al., ``Bootstrapped unsupervised sentence representation learning,'' in \textit{Proc. ACL-IJCNLP}, ACL, 2021, pp. 5168–5180.
%31
\bibitem{21Lee} D. Lee et al., ``Bootstrapping user and item representations for one-class collaborative filtering,'' in \textit{Proc. SIGIR}, ACL, 2021, pp. 5168–5180.
%32
\bibitem{06Vignat} C. Vignat et al., ``A geometric characterization of maximum r{\'e}nyi entropy distributions,”  in \textit{Proc. IEEE ISIT}, IEEE, 2006, pp. 1822–1826.
%33
\bibitem{20Wang} T. Wang and P. Isola, ``Understanding contrastive representation learning through alignment and uniformity on the hypersphere,'' in \textit{Proc. ICML}, PMLR, 2020, pp. 9929-9939.
%34
\bibitem{07Cohn} H. Cohn et al., ``Universally optimal distribution of points on spheres,'' \textit{Journal of the American Mathematical Society}, vol. 20, no. 1, pp. 99–148, 2007.
%35
\bibitem{19Borodachov} S. Borodachov et al., \textit{Discrete energy on rectifiable sets}, Springer, 1\textsuperscript{st} ed, New York, USA, 2019.
%36
\bibitem{15Vilnis} L. Vilnis et al., ``Word representations via gaussian embedding,'' in \textit{Proc. ICLR}, 2015.
%37
\bibitem{18Bojchevski} A. Bojchevski and S. G{\"u}nnemann, ``Deep gaussian embedding of graphs: Unsupervised inductive learning via ranking,” in \textit{Proc. ICLR}, 2018.
%38
\bibitem{19Oh} S. J. Oh et al., ``Modeling uncertainty with hedged instance embedding,'' in \textit{Proc. ICLR}, 2019.
%39
\bibitem{19Shi} Y. Shi and A. K. Jain, ``Probabilistic face embeddings,'' in \textit{Proc. ICCV}, IEEE, 2019, pp. 6902-6911.
%40
\bibitem{21Chen} K. Chen et al., ``Reliable probabilistic face embeddings in the wild,'' \textit{arXiv preprint arXiv:2102.04075}, 2021, pp. 6902-6911.
%41
\bibitem{19Paszke} A. Paszke et al., ``PyTorch: An imperative style, high-performance deep learning library,'' in \textit{Proc. NeurIPS}, 2019, pp. 8024–8035.
%42
\bibitem{15Snyder} D. Snyder et al., ``Musan: A music, speech, and noise corpus,'' \textit{arXiv preprint arXiv:1510.08484}, 2015.
%43
\bibitem{17Ko} T. Ko et al., ``A study on data augmentation of reverberant speech for robust speech recognition,'' in \textit{Proc. ICASSP}, IEEE, 2017.
%44
\bibitem{16Ulyanov} D. Ulyanov et al., ``Instance normalization: The missing ingredient for fast stylization,'' in \textit{Proc. CVPR}, IEEE, 2016.
%45
\bibitem{15Kingma} D.P. Kingma and L. J. Ba, ``Adam: A method for stochastic optimization,'' in \textit{Proc. ICLR}, 2015.
%46
\bibitem{20Lee} J. Lee et al., ``Momentum contrast speaker representation learning,'' \textit{arXiv preprint arXiv:2010.11457}, 2020.
%47
\bibitem{19Park} DS. Park et al., ``Specaugment: A simple data augmentation method for automatic speech recognition,'' in \textit{Proc. INTERSPEECH}, 2019, pp. 2613--2617.
%48
\bibitem{20Nagrani2} A. Nagrani et al., ``Disentangled speech embeddings using cross-modal self-supervision,'' in \textit{Proc. ICASSP}, IEEE, 2020.
%49
\bibitem{20Chung2} S.-W. Chung et al., ``Seeing voices and hearing voices:  learning discriminative embeddings using cross-modal self-supervision,'' in \textit{Proc. INTERSPEECH}, 2020.
%50
\bibitem{11Dehak} N. Dehak et al., ``Front-end factor analysis for speaker verification,'' \textit{IEEE Transactions on Audio, Speech, and Language Processing}, vol. 19, no. 4, pp. 788--798, 2010.
%51
\bibitem{06Ioffe} S. Ioffe et al., ``Probabilistic linear discriminant analysis,'' in \textit{Proc. ECCV}, Springer, 2006, pp. 531–542.
\end{thebibliography}
\end{document}